\documentclass{ws-ijmpa}
\usepackage[super,compress]{cite}
\usepackage{graphicx}
\usepackage{xcolor}
\usepackage{dsfont}
\usepackage{tikz}
\usepackage{tikz-cd}
\usetikzlibrary{scopes,intersections}
\usetikzlibrary{calc}
\usetikzlibrary{calc,decorations.pathmorphing,patterns}
\usepackage{pgfplots}
\usepackage{pst-solides3d}
\DeclareFontFamily{OT1}{pzc}{}
\DeclareFontShape{OT1}{pzc}{m}{it}{<-> s * [1.100] pzcmi7t}{}
\DeclareMathAlphabet{\mathpzc}{OT1}{pzc}{m}{it}
\DeclareMathAlphabet{\mathpzc}{OT1}{pzc}{m}{it}
\usepackage{bbm}
\usepackage{amssymb}
\usepackage{amsmath}
\usepackage{textcomp}

\usepackage{upgreek}
\usepackage{braket}
\UseRawInputEncoding
\begin{document}
\markboth{Slagter, R. J.}{New gravitational instanton}
\catchline{}{}{}{}{}
\title{On a quantum black hole}
\author{Reinoud Jan Slagter\footnote{info@asfyon.com}}
\address{Asfyon, Astronomisch Fysisch Onderzoek Nederland  \\and retired from\\
University of Amsterdam, ITP The Netherlands}
\maketitle
\begin{history}
\received{Day Month Year}
\revised{Day Month Year}
\end{history}
\begin{abstract}
We present an exact gravitational instanton solution on a five-dimensional, conformally invariant, Kerr-like warped Riemannian brane-world manifold. This geometry can be described as the K\"ahler manifold $\mathbb{C}^1\times\mathbb{C}^1\times \mathbb{R}$. By applying a double cover of $S^3$ through stereographic projection onto $\mathbb{C}P^1\times \mathbb{C}P^1$
of the effective four-dimensional manifold, together with the Klein surface construction, we exploit the underlying $\mathbb{Z}_2$ symmetry. The instanton is then obtained by fibering over the antipodal $S^2$.
The metric is determined by a first-order differential equation containing an integer parameter, while the equation governing the angular momentum component decouples from the remaining field equations. 
We demonstrate that our solution can be transformed analytically into a locally conformally related K\"ahler manifold with a K\"ahler potential, thereby revealing the self-dual structure. Finally, we explore a potential connection with Penrose's model of gravity-induced collapse in entangled primordial spacetimes. 
\centerline{\footnotesize Received: ;$\;$
Revised: ;$\;$ Accepted: .}\bigskip\bigskip\bigskip
\end{abstract}
\keywords{Instantons . Primordial black holes . Conformal invariance . Brane world models . Klein surface . Antipodicity . K\"ahler manifold . Self-duality . Icosahedron . Measurement problem}
\maketitle
\section{Introduction}\label{intro}
In this manuscript, we extend our previous work on an exact time-dependent instanton solution on a vacuum Kerr-like warped spacetime in conformal dilaton gravity, where the metric is decomposed as $g_{\mu\nu}=\omega^{4/(n-2)}\tilde g_{\mu\nu}$, 
with $\tilde g_{\mu\nu}$ representing the conformally related ('unphysical') metric and $\omega$ the dilaton (or gravitational scalar) field \cite{slagter2023,slagter2025b,slagter2026,slagter2026b}. Remarkably, the metric is determined by a first-order partial differential equation, providing a natural connection with self-duality (SD).
The singular points of the solution are determined by the roots of a quintic polynomial. Their locations are intimately related to the stereographic projection of the Riemann sphere onto the complex plane. During black-hole evaporation, these roots undergo transformations governed by the symmetry group of the icosahedron, namely the 60 rotational elements of the alternating group $A_5$. This observation suggests that a quintic polynomial represents the highest-order algebraic structure describing the singularities of axially symmetric Petrov type D black holes. Consequently, the solution lies outside the standard Pleba\'nski-Demia\'nski classification, which is characterized by a quartic structure function \cite{griff2005}.
The effective four-dimensional geometry can be described locally as a conformally K\"ahler manifold with Euclidean signature and an associated K\"ahler potential \cite{calabi1979,aksteiner2022,krasnov2020}. Its SD character makes comparison with the Yang--Mills instanton particularly appealing. Although a genuine five-dimensional K\"ahler manifold cannot exist, the effective four-dimensional geometry obtained on the brane possesses a SD K\"ahler structure with a recurrent conformal geometry. We conjecture that this effective geometry represents a gravitational instanton.
The construction arises through the projection of the five-dimensional Weyl tensor onto the brane \cite{ran1999,ran1999a,shir2000}. The resulting Kaluza Klein (KK) contribution encodes information about bulk gravitational waves and influences the evolution of the induced brane geometry. Should Hawking radiation eventually be observed with sufficient precision, it may provide indirect evidence for the presence of the extra bulk dimension.
We also employ the two-fold Riemannian covering associated with the Klein correspondence, relating the topology to $\mathbb{C}P^2$.
The topology of the gravitational instanton is taken to be $S^3\times\mathbb{R}/\mathbb{Z}_2$. An antipodal boundary condition is imposed on the hypersurface of a Klein bottle, $\sim\mathbb{C}^1\times\mathbb{C}^1$, to describe the Hawking particles produced during the evaporation process. By employing the Hopf fibration, the black hole horizon is represented by $S^2$, while its centrix is located not on a torus but on the Klein bottle. The intrinsic twist of the Klein bottle is naturally compatible with the antipodal identification of points on the horizon. Consequently, no 'cut-and-paste' construction is required \cite{thooft2016}, and the Hawking particles remain in a pure quantum state without the need for instantaneous transport. Instead, they propagate for a finite interval along the Klein surface.
We further speculate that the recently observed primordial black hole candidates, the so-called 'little red dots' in the very early Universe, may be related to primordial black holes created through instanton-mediated processes.
A further result of this work is the identification of a connection between the interior geometry of our black hole solution and the model proposed by Janis, Newman, and Winicour \cite{janis1968}, who described the Schwarzschild solution coupled to a massless scalar field, leading to an anomalous asymmetry. Moreover, our Kerr-like spinning solution bears an intriguing resemblance to the celebrated Newman-Janis construction \cite{janis1965}, in which the Kerr metric is generated from the Schwarzschild solution by means of a complex coordinate transformation and interpreted as the gravitational field of a rotating ring singularity.
Our gravitational instanton will be compared with two well-known examples, namely the Eguchi Hanson (EH) \cite{eguchi1979} and the Fubini Study (FS) \cite{krasnov2020} gravitational instantons. The present framework can be extended straightforwardly to non-vacuum configurations by incorporating an additional scalar field \cite{slagter2023}. Near the Planck scale, both the dilaton (or gravi-scalar) and the scalar field may be regarded as quantum fields.
The remainder of this paper is organized as follows. Section 2 summarizes the underlying conformally invariant model. Section 3 reviews the two best-known locally conformal K\"ahler gravitational instantons, namely the EH and FS solutions. We also discuss the Newman Janis complex transformation of the Schwarzschild metric and its interpretation of the Schwarzschild singularity, since these constructions provide important motivation for our approach. Section 4  presents the new gravitational instanton and investigate its geometrical and physical properties.
\section{Summary of the model}\label{sec:2}
Recently we studied\cite{slagter2023,slagter2025b,slagter2026,slagter2026b}, on the Randall Sundrum (RS) brane world model\cite{ran1999,ran1999a,shir2000} with  a conformal invariant Lagrangian, a new exact Kerr-like black hole solution. The Lagrangian is given by
\begin{eqnarray}
S=\int d^dx\sqrt{-g}\Bigl[-\frac{1}{2}\xi (\Phi\Phi^*+\omega^2)R
-\frac{1}{2}\tilde g^{\mu\nu} \Bigl({\cal D}_\mu\Phi({\cal D}_\nu\Phi)^*+\partial_\mu\omega\partial_\nu\omega\Bigr)\qquad \cr
-\frac{1}{4}F_{\mu\nu}F^{\mu\nu}-V(\Phi,\omega)-\Lambda\kappa^{\frac{4}{d-2}}\xi^{\frac{d}{d-2}}\omega^{\frac{2d}{d-2}}\Bigr],\qquad\label{2.1}
\end{eqnarray}
with $\Phi$ a scalar field, $\omega$ the phantom (dilaton) field, $F$  the Abelian field strength, $V(\Phi,\omega)$ the potential, $\Lambda$ the cosmological constant and  $\xi=(d-2)/(4(d-1))$. Further, ${\cal D}$  is the gauge covariant derivative.
We applied 't Hooft's trick\cite{thooft2016,thooft2019,thooft2021,thooft2015}, by writing (for the case $d=5$) $^{(5)}{g_{\mu\nu}}=\omega^{4/3} {^{(5)}{\tilde g_{\mu\nu}}}$ and    ${^{(5)}{\tilde g_{\mu\nu}}}=^{(4)}{\tilde g_{\mu\nu}}+n_\mu n_\nu$. An 'unphysical' spacetime is thus  separated. Here $n_\mu$ is the unit normal to the brane. Again we write $^{(4)}{\tilde g_{\mu\nu}}=\bar\omega^2{^{(4)}{\bar g_{\mu\nu}}}$.
We redefined $\omega^2\rightarrow -6\frac{\omega^2}{\kappa^2}$, with $\kappa=8\pi G_N$, to ensure that the field $\omega$ has the same unitarity and positivity properties as the scalar field $\Phi$. 
The $\omega$  is used by the in-falling observer to describe his experience of the vacuum.
The Lagrangian is conformally invariant, i.e.,
$g_{\mu\nu}\rightarrow  \Omega^{4/(d-2)}g_{\mu\nu}, \omega\rightarrow \Omega^{(2-d)/2}\omega, 
\Phi\rightarrow \Omega^{(2-d)/2}$.
This gauge freedom can be employed to describe the different experiences of local and distant observers when studying Hawking evaporation, while simultaneously allowing the manifold to be rendered conformally K\"ahler. The conformal symmetry may be spontaneously broken by introducing a mass term in $V(\omega,\Phi)$, analogous to the Higgs mechanism.
The principle of black hole complementarity can then be invoked: an observer falling into a black hole may describe physical processes in a fundamentally different way from an external observer, with these apparent discrepancies potentially being related through conformal transformations \cite{thooft2009,thooft2010}. It is conceivable that black holes, spacetime singularities, and event horizons are not fundamental structures, but rather emergent features that disappear in a more complete description, while causality and locality remain essential principles of quantum gravity.
Furthermore, black holes might represent large-scale limits of more regular field configurations that become apparent at smaller length scales. In this context, a gravitational instanton may provide a possible candidate for such an underlying regular configuration.
We consider the 5D metric
\begin{eqnarray}
ds^2=\omega(t,r,y_5)^{4/3}y_0\Bigl[-N(t,r)^2dt^2+\frac{1}{N(t,r)^2}dr^2+dz^2\cr +r^2(d\varphi+N^\varphi(t,r)dt)^2+dy_5^2\Bigr],\qquad\quad\label{2.3} 
\end{eqnarray}
where $y_5$ is the bulk coordinate and $y_0$ the bulk dimension. 
When applied in a cosmological setting, $\omega$ becomes a  'warp' factor and  will be, in the most simple situation, of the form $\sim e^{\sqrt{\Lambda_5/(y_5-y_0)}}$. In general, the $(r,t)$ dependent part can be solved exactly in the FLRW case\cite{slagter2016}.
We switch to Euclidean space by performing the Wick rotation $t \rightarrow i\tau$ together with $N^\varphi \rightarrow iN^\varphi$.
Since we are working within a RS warped brane-world model, the five-dimensional bulk Einstein equations and the effective four-dimensional brane equations must be solved simultaneously. The latter contain the contribution of the projected bulk Weyl tensor, which encodes the non-local gravitational effects of the bulk geometry on the brane. Consequently, the complete system becomes a coupled set of five- and four-dimensional field equations,
\begin{equation}
\omega^2{^{(5)}}{G_{\mu\nu}}-{^{(5)}}{T^{(\omega)}_{\mu\nu}}
+\frac{3}{4\sqrt[3]{12}}\Lambda_5\kappa_5^{4/3}\omega^{10/3}{^{(5)}}{g_{\mu\nu}}=0,\label{2.4}
\end{equation}
\begin{equation}
{^{(4)}}{G_{\mu\nu}}-\frac{1}{\omega^2}\Bigl[{^{(4)}}{T^{(\omega)}_{\mu\nu}}-\frac{1}{6}\Lambda_{eff}\kappa_4^2\omega^4{^{(4)}}{g_{\mu\nu}}\Bigr]+{\cal E}_{\mu\nu}=0.\label{2.5}
\end{equation}
where $T_{\mu\nu}^{(\omega)}$  is the contribution from the dilaton (which appears in both equations)
\begin{eqnarray}
T_{\mu\nu}^{(\omega)}=\nabla_\mu\nabla_\nu\omega^2- g_{\mu\nu}\nabla^2\omega^2
+\frac{1}{\xi}\Bigl(\frac{1}{2} g_{\alpha\beta} g_{\mu\nu}- g_{\mu\alpha} g_{\nu\beta}\Bigr)\partial^\alpha\omega\partial^\beta\omega.\label{2.6}
\end{eqnarray}
and  ${\cal E}_{\mu\nu}$ the contribution from the bulk.
We are  mainly interested in the stationary case. It turns out that the equations for the bulk, as well as for the effective brane are
\begin{equation}
\ddot\omega=-N^4\omega''+\frac{d}{\omega(d-2)}\Bigl(N^4\omega'^2+\dot\omega^2\Bigr),\label{2.7}
\end{equation}
\begin{eqnarray}
\ddot N=\frac{3\dot N^2}{N}-N^4\Bigl(N''+\frac{3N'}{r}+\frac{N'^2}{N}\Bigr)\qquad\qquad\qquad\cr
-\frac{(d-1)N^5}{(d-3)\omega}\Bigl[\omega''+\frac{\omega'}{r}+\frac{d}{2-d}\frac{{\omega'}^2}{\omega}\Bigl)+\frac{\omega' N'}{N}+\frac{\dot\omega\dot N}{N^5}\Bigr].\label{2.8}
\end{eqnarray}
So the spin part decouples.
The exact solution becomes
\begin{eqnarray}
N^2=\frac{C_2}{r^2((t-t_0)^4+C_3)}\Bigl[\frac{(r-a)^{k+1}\Bigl(r(k+1)+a\Bigr)}{k+2}+C_1 \Bigr],
\cr\omega = \Bigl(\frac{b_i}{(r-a)(t-t_0)}\Bigr)^{\frac{1}{2}d-1}, N^\varphi=\int \frac{dr}{r^3\omega^{\frac{d-1}{d-3}}}+F_d(t),\quad\label{2.9}
\end{eqnarray}
\begin{figure}
\centerline{
	\includegraphics[width=3.7cm]{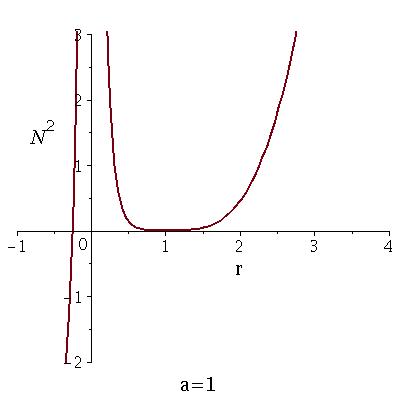}
	\includegraphics[width=3.7cm]{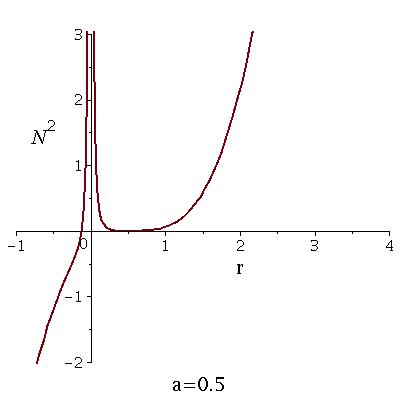}}
	\centerline{
	\includegraphics[width=3.7cm]{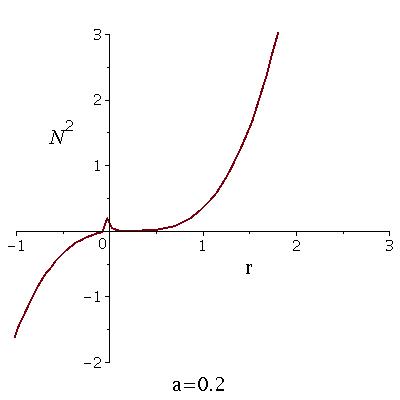}
	\includegraphics[width=3.7cm]{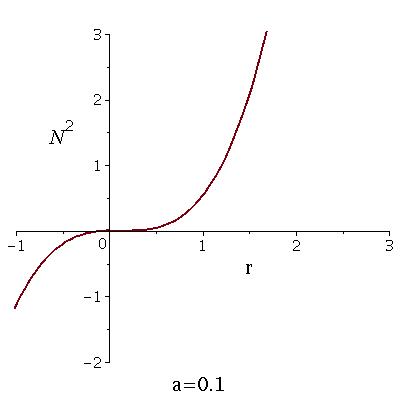}}	 	
	\caption{{\it Plot of $N(r)^2$ for several values of $a$. We observe for large value of a, there are two zero's. When a decreases, it turns out that the for $r\rightarrow \pm 0$, $N^2$ tends to $+ \infty$. Finally the singularity at $r =0$  does not exist there. Reversing this evolution suggests that black-hole horizons may emerge from a regular gravitational instanton configuration. After complexifying our manifold, we obtain, as we shall see, a markedly different behavior.}} 
	\label{fig1}
\end{figure}
with constants $(a,b_i,t_0,C_i)$ and $k$ an integer and $F_d(t)$ an arbitrary function. Note that the solution for $N$ is the same for the bulk and brane manifold, as it should be (apart from the values of the constants). In Fig.(\ref{fig1}) we plotted its behavior. The dilaton depends on the dimension. 
It is remarkable that the r-dependent part $N(r)$, is determined by a first order differential equation,
\begin{equation}
rN\frac{\partial N}{\partial r}+N^2=\frac{k+1}{2}(r-a)^k.\label{2.10}
\end{equation}
\begin{figure}
\centerline{
\includegraphics[width=4cm]{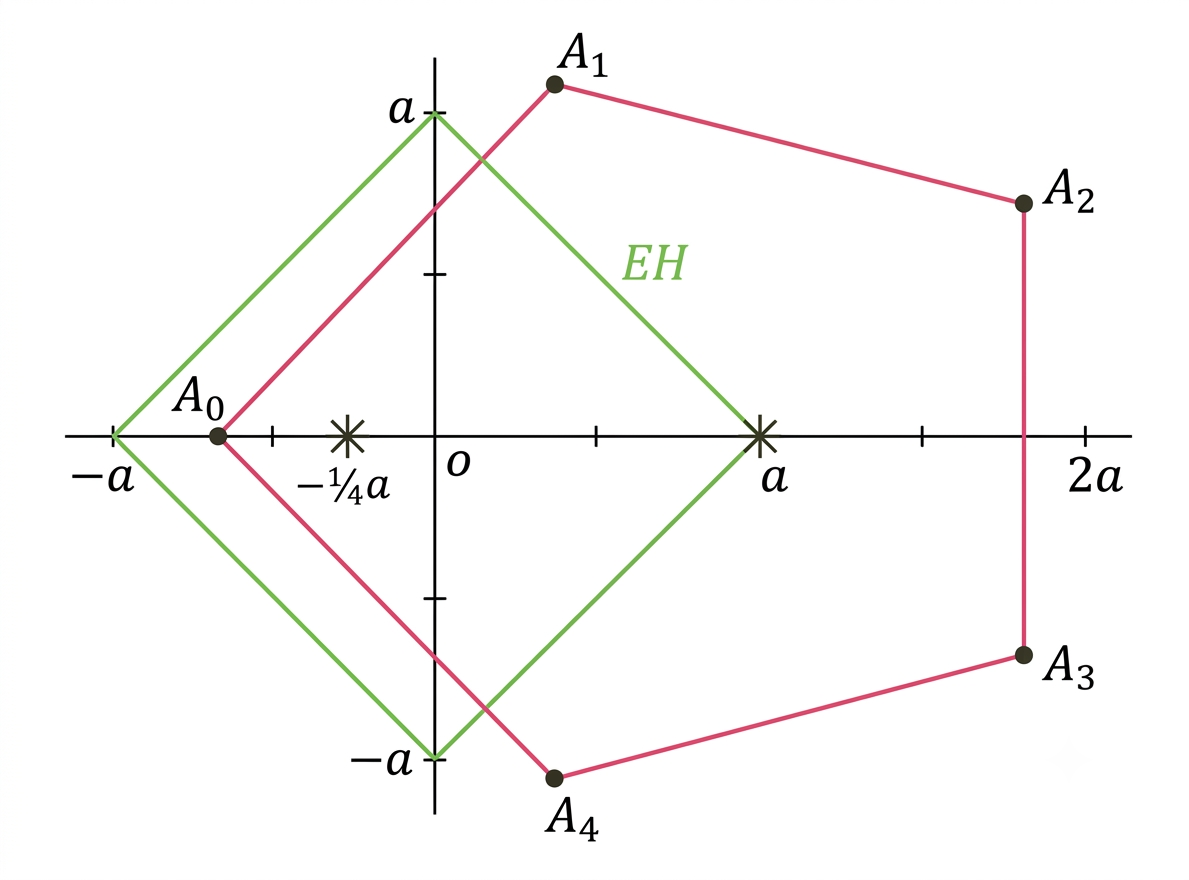}
\includegraphics[width=4.cm]{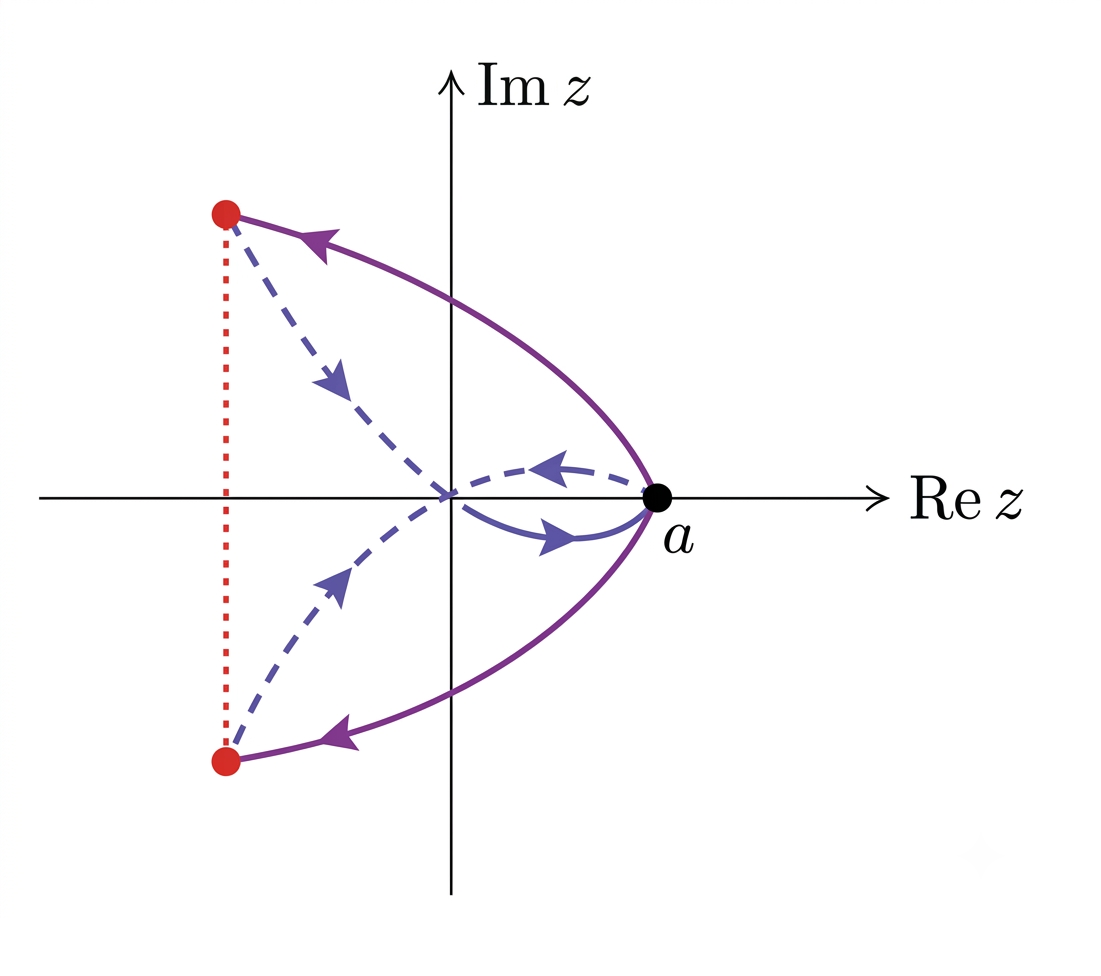}}
\caption{ Left: Possible locations of the zeros of the quintic (red) compared with those of the Eguchi-Hanson  case (green). Right: Image of $z=a$ in the complex plane. During black hole evaporation, the two complex zeros move toward $z=-a$, while the geometry remains free of singular behavior.}
\label{fig2}
\end{figure}
In Fig.(\ref{fig2}) and Fig.(\ref{fig3}) we plotted the behavior of the roots in the complex plane.
For $k=3$, $N$ is a quintic polynomial, i.e., $4r^5-15ar^4+20a^2r^3-10a^3r^2+C_1=-r^5+5r^4(r-a)-10r^3(r-a)^2+10r^2(r-a)^3+C_1$, with in general five zero's, which can become complex.
For $C_1=0$ the zeros are at $r_H=a$ with multiplicity $(k+1)$ and at $r_H=-\frac{a}{k+1}$. 
For $k=3$ and varying $C_1$, one obtains a distribution of the zero's which are
related to  the stereographic projection of the icosahedron.  The general solution of the quintic can be found by means of elliptic curves.
For $k=2$ and $C_1 < a^4/4$ there are 4 zeros, as in the Banodoz--Teitelboim-Zamelli ( BTZ)  solution, two of them negative. 
For $k=1$ and $C_1 < -a^3/3$ there is still a real solution.
For $k=0$ we  surprisingly get two horizons $r_H=\pm a$. This solution is also found from the field equations for constant $\omega$. The differential equation for $N$ becomes then  $N''+3N'/r+N'^2/N=0$, with solution $N\sim(1-a^2/r^2)$.
One proves that in this case a complex 2-dimensional manifold M endowed with metric g, admits a local K\"ahler immersion into a Hilbert space, i.e., a real 3-dimensional K\"ahler manifold N, if g is a real analytic K\"ahler metric. The immersion, $f:M\rightarrow N,$ is the given by $ (z_1,z_2)\rightarrow (z_1,z_2,f(z))$, with f holomorphic with no terms of degree less than 2\cite{calabi1979}.
We will use this property.
Because we must solve the 5D and 4D equations simultaneously, we get two constraint equations with the cosmological constants $\Lambda_5$ and $\Lambda_{eff}$. These equations are fulfilled, if one allows a 'fine tuning', 
\begin{equation}
\Lambda_{eff}=\frac{3}{8}\sqrt[3]{18}\frac{\kappa_5^{4/3}b_5^2}{\kappa_4^2b_4^2}\Lambda_5,\quad C_4=\frac{C_5b_4^3}{b_5^2},\label{2.13} 
\end{equation}
with the constants $(C_4,b_4)$ and $(C_5,b_5)$ belonging to the two solutions of the angular momentum and $\omega$ respectively.
This proves once again that the 5D and 4D effective equations must be solved together. The situation changes, when we incorporate a scalar field. 
Cauchy's theorem on holomorphic functions $F(z)$ on an open set of the complex plane, tells us, within a simply-connected region with no singularities inside a loop C, that
$\oint_C F(z)dz=0.$
Our solution for $N$ can also be written as
\begin{equation}
N^2=\frac{4}{z^2}\int z(z-a)^3dz=0.\label{2.15}
\end{equation}
So we have $F(z)=z(z-a)^3$.
One applies  Cauchy's integral theorem
\begin{equation}
F(z)=\frac{1}{2\pi i}\oint_C\frac{F(\xi)}{\xi -z} d\xi,\label{2.16}
\end{equation}
\begin{figure}[h]
	\centerline{
	\includegraphics[width=3.3cm]{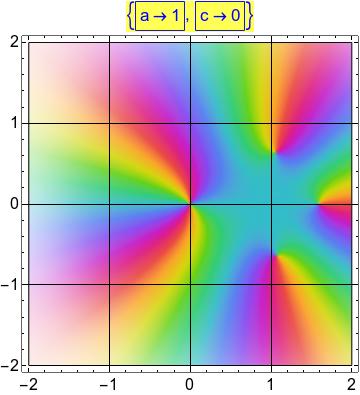}
	\includegraphics[width=3.4cm]{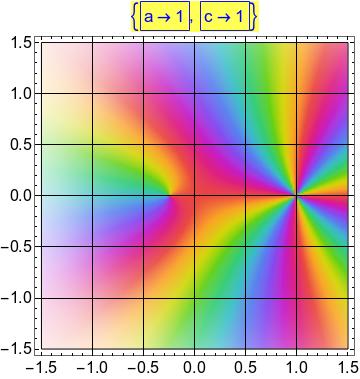}}
	\centerline{
	\includegraphics[width=3.cm]{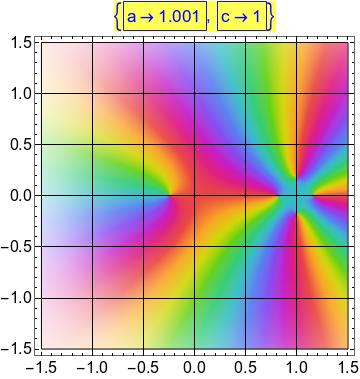}		
	\includegraphics[width=3.cm]{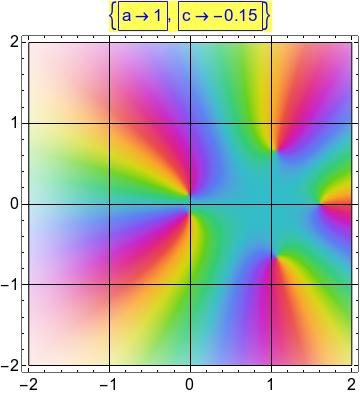}
	\includegraphics[width=3.cm]{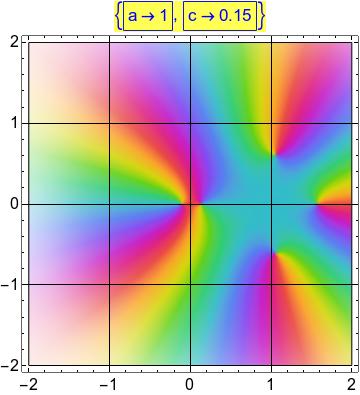}}	
	\caption{{\it 'Dance' of the roots in the complex plane for several values of $a$ and $C_1$.}} 
	\label{fig3}
\end{figure}
with z inside C. 
The values of an analytic function are uniquely determined by its values on the boundary. In the present construction, the points $z=(0,a)$ appear as removable singularities. Since they lie on the real axis, their contribution can be treated by evaluating the corresponding Cauchy principal values.
Under the antipodal identification, no fixed points are present. Consequently, the quintic polynomial determines the locations of the singular points on $S^3$ within the complex manifold C, without introducing additional pathologies when the C shrinks to a point. Because of the extra bulk dimension, the deformation of C does not intersect the singular loci.
Nevertheless, the conformal $\Omega$ freedom remains available. This gauge freedom can be used to choose a conformal frame in which the asymptotic observer experiences an effective Kerr-like black hole geometry. Finally, we should mention that there is an interesting connection between the geometry of polyhedra (in our case, the tetrahedron and the icosahedron) and loop quantum gravity\cite{bianchi2017}.
Any finite subgroup of $S^3 = SU(2)$ is either cyclic, or conjugate to one of the subgroups $(D_d^*, T^*, O^*, I^*)$\cite{toth2002}.
One proves that when a manifold is a fibered solid Klein bottle, then M is double covered by a trivial fibered solid torus $S^2\times D^2$.
The corresponding action of $\mathbb{Z}_2$ on  $S^1\times D^2$ is generated by a homeomorphism which is simply the product of a rotation through $\pi$ on the
$S^1$-factor with a reflection of the $D^2$-factor. Here $D^2$ represents the binary dihedral group $D^2=\{ e^{i\pi l/d}, l=0 ... 2d-1\}\cup j.\{ e^{i\pi l/d}, l=0 ... 2d-1\}$.
One describes the finite M\"obius groups by the skew-field of quaternions $H=\mathbb{R}^4$ with canonical basis $(i,i,j,k)$. The group of unit quaternions $S^3 \subset H$ is isomorphic with $SU(2)$. One can visualize these groups by the Clifford decomposition of $S^3$, parametrized by quaternion $z+iw$, with $|z|^2+|w|^2=1$ (section (6.3)). The level set is
$T_t=\{(z,w)\in S^3\subset\mathbb{C}^2, |z|^2-|w|^2=\cos(2t)\},\quad z=\cos(t) e^{i\theta}, w=-\sin(t) e^{i\varphi}$
with $0\leq t\leq \pi/2, 0\leq (\theta ,\varphi) <2\pi$.  For $t=0, \pi/2$ we have two circles $T_0, T_{\pi/2}$ cut out from $S^3$. We obtain so the Clifford Klein bottle. The binary groups  fit these decomposition.
The $S^3$ has the structure of a non-commutative group, with center $(\pm 1)$ The quotient $S^3/\{\pm 1\}$ is isomorphic to  $SO(3)$ of orientation preserving isometries of $S^2$.
\begin{figure}[h]
\centerline{
\includegraphics[width=0.4\textwidth]{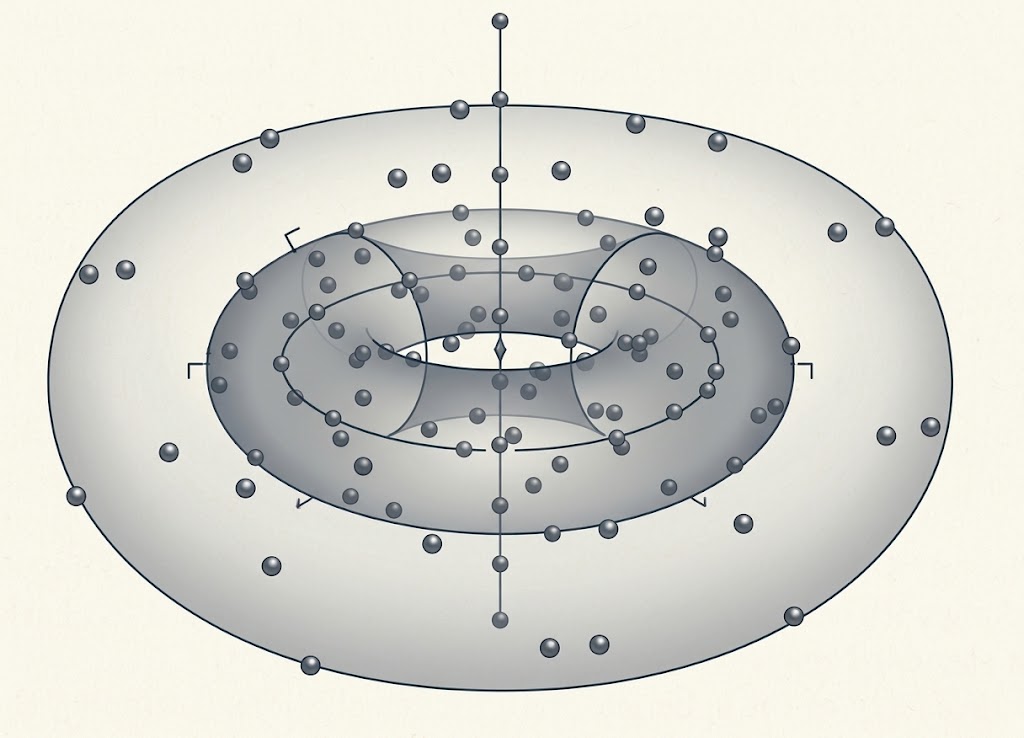}}
\centerline{
\includegraphics[width=0.4\textwidth]{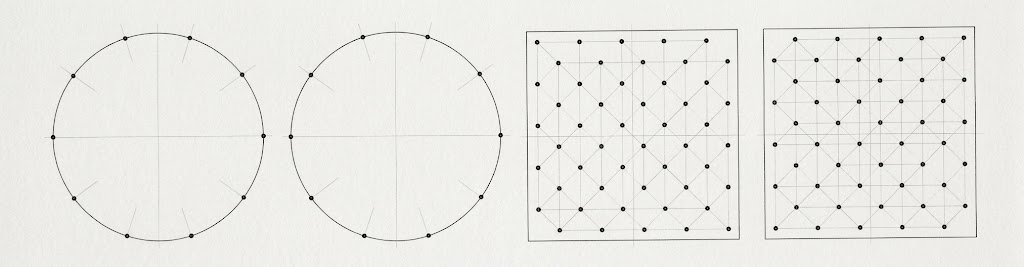}}
\caption{{\it  A 3-dimensional view of the binary icosahedral group $I^*$\cite{toth2002}.}}\label{icogroup}
\end{figure}
The icosahedral group $I^*$ possesses 120 elements. See Fig.(\ref{icogroup}), where we plotted the elements for the torus case ($w=\sin(t)e^{i\varphi}$).
The vertices of the icosahedron stereographically projected on $\mathbb{C}$ are $(0, \infty, \omega+\omega^4),\omega^j(\omega^2+\omega^3), j=0...4, \omega =e^{2\pi i/5}$.
$A_n$ is the commutator subgroup of the symmetric group $S^n$ with index 2 and has therefore $n!/2$ elements. It is the kernel of the signature group homomorphism,
explained under symmetric group. We were interested in the alternating group of the icosahedron with $5! /2 = 60$ elements, equivalent with the 5 orbits or symmetry rotations, times 12. One can consider these 12 elements of each orbit  as entangled symmetry (qubit)\cite{Aulbach2011}.
It are the 5 regular tetrahedra circumscribed around the icosahedron. An orbit  delivers the vertices  of a pentagon.
We conjecture that the $60$ elements, are the ontological states for our black hole. 
Any finite subgroup of $S^3=SU(2)$ is either cyclic, or conjugate to one of the binary subgroups mentioned above.
The invariant forms\cite{toth2002} of the tetrahedron  and icosahedron, play a fundamental role in describing the singular points in our black hole model.
The vertices of the tetrahedron are the antipodals of the original tetrahedron. The midpoints of the edges of any of the tetrahedron are the vertices of the octahedron. Projected at $\hat{ \mathbb{C}}$, these are $(0,\infty,\pm1,\pm i)$.
A different route could be provide by linking quantum features to the representation of tetrahedra and the $SU(2)$. A Hilbert space of the 'quantum' tetrahedra should results\cite{barbieri1998}. Even could one formulate symmetric states and entangled states. A central tool in this method is the Majorana representation, a generalization of the Bloch sphere. 
It is a method for  geometrical measurement of entanglement.

Any n-qubit state can be visualized by a multi set of n points on the surface of $S^2$.
Under the local unitary operations on subsystems, the Majorana points (MP) rotate rigidly and the unitary operations $U\in SU(2)$ on a single qubit, resulting in a rotation of the Bloch sphere.
Any symmetric state of n qubits $\ket{\Phi^s}$, can be composed, up to a phase, from a sum over all permutations $G\in S_n$ of n indistinguishable single qubits states $[\ket{\phi_1}...\ket{\phi_n}]$, with $S_n$ the symmetry group. It turns out that there is a direct relationship between the symmetric and non-symmetric physical states and the MP's, with the complex polynomials belonging to stereographic projections of Platonic tilings, circumscribed (or inscribed) the Riemann sphere.
Let us consider the (Majorana) polynomial for a symmetric state $\ket{\Phi^s}=\sum_{k=0}^n a_n\ket{S_{n,k}}$,
\begin{equation}
\Phi(z)=\sum_{k=0}^n (-1)^k\sqrt{\binom{n}{k}} a_k z^k=A\prod_{k=1}^n (z-z_k).\label{2-17}
\end{equation}
An MP at the North Pole also corresponds to a point at infinity here. Strictly speaking, therefore, in Eq.\ref{2-17} we must take the sum to $n-r$, where r is the number of MP's being $\ket{0}$.
The connection with the singular points in our black hole model is evident. We could write our polynomial as 
\begin{equation}
N^2(z)=z^5+\frac{3}{4}\sqrt5 az^4+\frac{1}{2}\sqrt{10}a^2z^3+\frac{1}{4}\sqrt{10}a^3z^2+C.\label{2-18}
\end{equation}
However, the general quintic  is hard  to solve  by radicals.
\begin{figure}[h]
\centerline{
\includegraphics[width=0.45\textwidth]{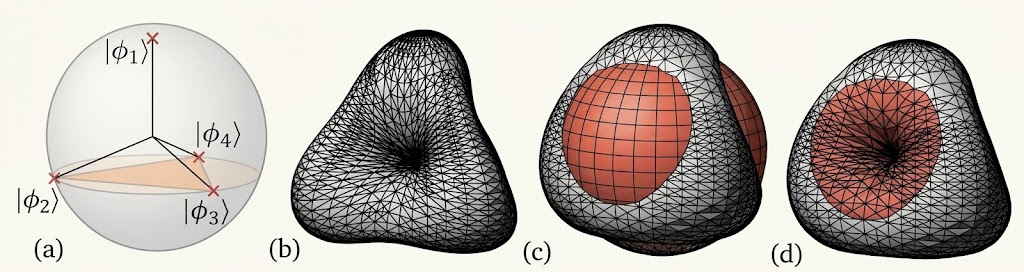}}
\centerline{
\includegraphics[width=0.45\textwidth]{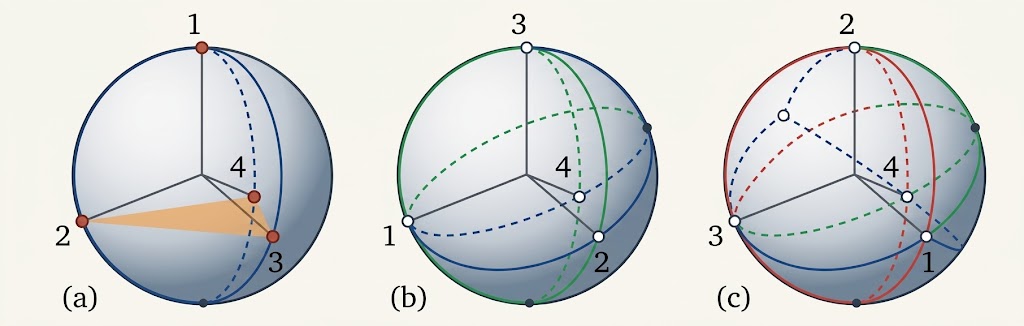}}
\centerline{
\includegraphics[width=0.45\textwidth]{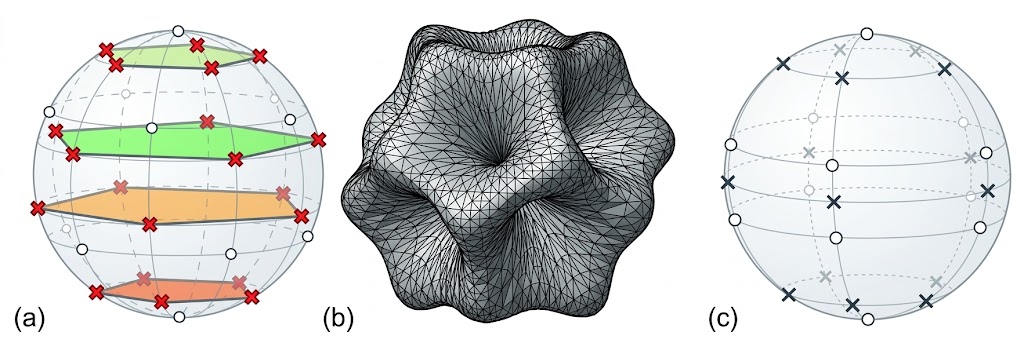}}
\caption{{\it Representations of the $n=4$ (top) and $n=12$ (bottom) qubits. Top left:   Visualisation of the tetrahedron state with Majorana points (MP) and closest product points (CPP), with one at each vertex pictured in (a). (b) shows the spherical amplitude functions and (c) and (d) show the volume functions. See text. 
Top right: The four CPP's. The MP's are determined by rotations of the CPP's according to the tetrahedral symmetry group. These permutations restrict the locations of the CPP's. The result is shown in (c).
Bottom: The 12-qubit $\ket{\Phi_{12}}$  icosahedron state (circular dots) and the 20 CPP's (red crosses), i.e. 10 with their antipodal counterparts. These correspond to the centers of the 20 faces\cite{Aulbach2011}.}}\label{majo2}
\end{figure}
\section{Local conformal K\"ahler manifolds}\label{sec:3}
The possible transition between the Schwarzschild and Kerr solutions, and vice versa, was discussed long ago by Israel and Penrose in connection with the nature of the central point singularity. Israel's theorem establishes that the Schwarzschild solution is the unique static, asymptotically flat vacuum solution possessing a regular event horizon. In contrast, an asymmetric collapsing body is expected to radiate away its higher multipole moments before settling into the Schwarzschild geometry. If the event horizon itself were singular, the formation of trapped surfaces would be obstructed, preventing matter from crossing the horizon. In such a scenario, the assumptions underlying Penrose's singularity theorem would no longer apply, suggesting that the conventional picture of gravitational collapse requires reconsideration.
Subsequently, Janis {\it et al.} derived an exact Schwarzschild-like solution in the presence of a massless scalar field. In this solution, sometimes referred to as the 'truncated' Schwarzschild solution, the Schwarzschild horizon is replaced by a point singularity. They further conjectured that every asymptotically flat, static solution possessing both mass and asymmetry together with a point singularity belongs to this class. If this conjecture is correct, then the standard description of collapse beyond the Schwarzschild radius must be revised, since the black hole interior would necessarily possess a geometry different from that of the classical Schwarzschild solution.
A similar conclusion is reached in our model, where an antipodal boundary condition is imposed. In this framework, a classical central singularity cannot be regarded as physically sustainable without incorporating quantum effects. Consequently, a consistent description of the black-hole interior requires a quantum treatment, naturally leading to the consideration of gravitational instantons.
An additional motivation arises from the close relationship between Einstein and K\"ahler metrics, which are conformally related under suitable conditions. The K\"ahler form provides a natural route to self-duality, while exact SD instanton solutions are well known in Euclidean Yang Mills (YM) theory. A notable example is the Euclidean $SU(2)$ YM instanton, which has finite action, a SD field strength $F_{\mu\nu}$ localized near $r=0$, decays as $1/r^4$, and approaches a pure gauge configuration for the gauge potential $A_\mu$ at spatial infinity.
The use of complex coordinates is also well established in quantum mechanics. For example, both pure and mixed states of a two-level quantum system can be represented in the complex Hilbert space $\mathbb{C}^2$, where a state is written as $\ket{z}=(z_1,z_2)$. It is therefore natural to investigate whether four-dimensional space-times in general relativity may likewise admit an efficient description in terms of complex coordinates. One of the earliest applications of this idea was due to Ernst \cite{ernst1968}, who introduced a complex potential that greatly simplified the Einstein equations for stationary axisymmetric space-times. The resulting formalism has proved particularly useful for generating new exact solutions and for constructing matchings between black hole interior and exterior geometries.
Complex structures also arise naturally in the formalism of Dirac spinors in $(3+1)$ dimensions, where discrete CPT transformations correspond to simple operations on $SU(2,\mathbb{C})$ spinors. Furthermore, several Riemannian symmetric spaces admit elegant descriptions in terms of complex projective geometry. In particular, the complex projective space $\mathbb{C}P^2$ plays a prominent role as a compact K\"ahler manifold with the FS metric. Closely related projective constructions associated with the torus and the Klein bottle are of particular interest in our model.
Motivated by these observations, we consider the diffeomorphism $\phi:S^3\times\mathbb{R}\rightarrow\mathbb{C}^2\setminus\{0\}$,
which provides a natural complex parametrization of the underlying manifold. It is well known that on an oriented Riemannian four-manifold, a SD metric is characterized by a SD Weyl conformal curvature tensor. Classical examples include the round metric on $S^4$ and the FS metric on complex projective space. These geometries provide important prototypes for the SD structures that will be investigated in the following sections.
\subsection{The Eguchi-Hanson (EH) and Fubini-Study (FS) instantons}\label{3-1}
It soon became apparent that the complexification of a manifold provides significant advantages in the study of gravitational instantons , analogous to those encountered in Yang Mills theory. Eguchi and Hanson constructed the first explicit anti-self-dual (ASD) solution of the Euclidean Einstein equations, representing a gravitational instanton, on an axially symmetric manifold \cite{eguchi1979,flah1976,eguchi1980}.
\begin{equation}
ds^2_{EH}=\frac{1}{f(r)^2}dr^2+\frac{r^2}{4}\Bigl(\sigma_x^2+\sigma_y^2+f(r)^2\sigma_z^2\Bigr),\label{3.1}
\end{equation}
with $\sigma_i$ the basis one forms. The solution  $f^2=r^4/(r^4-a^4)$ is determined by a first order differential equation. 
When written in Euler angles $(\psi,\theta,\varphi)$, one of the angles must run from $(0,2\pi)$ in stead of $(0,4\pi)$, in order to remove  the $r=0$ singularity.
Defining $\rho^4=r^4-a^4$, the metric is becomes geodesically complete  without true singularities. One writes the manifold in K\"ahler form on $\mathbb{C}^2/\{0\}$
\begin{eqnarray}
K=\frac{\rho^2}{\sqrt{\rho^4+a^4}}(dz_1d\bar z_1+dz_2 d\bar z_2)+\frac{a^4}{\sqrt{\rho^4+a^4}}\partial\bar\partial\log\rho^2\cr =\partial\bar\partial\log\Bigl
(\frac{\rho^2e^{\sqrt{\rho^4+a^4}}}{a^2+\sqrt{\rho^4+a^4}}\Bigr),\qquad\qquad\qquad\label{3.2}
\end{eqnarray}
with $(z_1,z_2)=(x+iy,z+it)$ and $ \rho=\sqrt{z_i\bar z_i}$.
The singular point $\rho =0$ is removed by the antipodal identification $(z_1,z_2)\rightarrow (-z_1,-z_2)$.
Near $r=a$ the manifold is $\mathbb{R}^2\times S^2$. At each point of the two sphere $(\theta,\psi)$ there is attached an $\mathbb{R}^2$ which shrinks to a point for $r\rightarrow a$.
The manifold is  homotopic to $S^2$. For large r, the metric approaches a flat metric and the constant-$r$ hypersurfaces  are 3-spheres, with antipodal points identified. 
The boundary is the $SO(3)\sim \mathbb{R}P^3$, the group manifold for which $S^3\sim SU(2)$ is the double cover. The EH is locally Euclidean, i.e.,  $\mathbb{R}P^3 \sim S^3/\mathbb{Z}_2$ but not globally\cite{eguchi1979}.
The FS instanton is another very interesting compact Einstein SD or ASD manifold, isomorphic to the Euclidean 4-sphere or the complex projected plane $\mathbb{C}P^1\#\mathbb{C}P^1$. 
There exist several methods to prove self-duality. Well-known approaches include the Pleba\'nski chiral formulation \cite{krasnov2020} and the Calabi construction \cite{calabi1979}. By requiring the curvature tensor to be SD as a two-form, one obtains a second-order differential equation for the metric components. This equation generally contains two integration constants, one of which may be identified with the cosmological constant. However, a first-order formulation is preferable, as demonstrated by the EH manifold, where self-duality reduces the problem to a more tractable first-order equation.
Let us now consider the FS manifold, which provides another important example of a SD K\"ahler geometry.
\begin{equation}
ds^2_{FS}=\frac{1}{S(r)^2}\Bigl(dr^2+\frac{r^2}{4}\sigma_x^2+\frac{r^2S(r)}{4}\sigma_y^2+\sigma_z^2\Bigr),\label{3.3}
\end{equation}
with $S(r)=1+\alpha r^2$.
One needs, however, a term $\alpha g_{\mu\nu}$ on the right hand side of the Einstein equations. This appears to be a cosmological constant.
The FS is K\"ahler, i.e., of the form ($i=(1,2)$)
\begin{equation}
ds^2_{FS}=\frac{\partial^2\upkappa}{\partial z^i\partial \bar z^{\bar i}}d z^i d\bar z^{\bar i},\quad \upkappa =\frac{1}{\alpha}\log\Bigl(1+\alpha z^i\bar z^{\bar i}\Bigl),\label{3.4}
\end{equation}
with $\upkappa$ the K\"ahler potential. The metric is local conformal to
\begin{eqnarray}
ds^2_{FS}=(1-|z^1|^2)(dz^1 d\bar z^1+dz^2d\bar z^2)
 -\bar z^1 z^2d z^1d\bar z^2 - z^1 \bar z^2d \bar z^1d z^2.\label{3.5}
\end{eqnarray}
There exists a deep relation between the immersion (or embedding) of a real Riemannian K\"ahler manifold and its realization as a complex manifold. A fundamental result due to Calabi states that a complex manifold (M), equipped with a real analytic K\"ahler metric (g), admits a local K\"ahler immersion into a Hilbert space. This theorem establishes a direct connection between the intrinsic geometry of the manifold and its extrinsic complex embedding properties.
For manifolds with non-orientable structures, an additional stereographic projection or double-cover construction is required in order to consistently define the complex structure. Using the theory of fiber bundles together with Cartan's method of connections, Calabi derived a first-order differential equation governing the existence of such K\"ahler immersions. This formulation is particularly useful because, unlike the second-order curvature conditions obtained by imposing self-duality directly, it reduces the problem to a first-order system closely related to the underlying geometric structure. One obtains
\begin{equation}
u'=\frac{\pm\sqrt{1+4\alpha r}-1}{2\alpha r},\label{3.6}
\end{equation}
with solution
\begin{equation}
u(r)=\frac{1}{2\alpha}\Bigl(\beta+2\sqrt{1+4\alpha r}-2\ln[\sqrt{1+4\alpha r}+1]\Bigr),\label{3.7}
\end{equation}
with $(\alpha ,\beta)$ a constant. Remark that there is a striking similarity with our first order differential equation Eq. (\ref{2.10}).
We are also interested in conformal-K\"ahler forms, because our starting point is a conformal Lagrangian. The two-form $K$ can then be written as $\hat K=\Omega^2  K$, with $d\hat K=0$. Locally, one can write $\hat g_{ij}=\Omega^2 g_{i\tilde j}=\partial_{z^i}\partial_{\bar z^j} \upkappa$. 
Let us consider the block form with (almost) Hermitian structure and 4 complex scalars\cite{aksteiner2022,ornea2024,aksteiner2022b}
\begin{equation}
ds^2=g_{a\tilde b}dz^a d\tilde z^b =a_{ij}d\sigma^i d\sigma^j+b_{IJ}dx^I dx^J,\label{3.8}
\end{equation}
with $\sigma^i=(\tau,\varphi),\quad  x^I=(x,y)$
and with the two-form $K=ig_{a\tilde b}(z^a, \tilde z^b)dz^a\wedge d\bar z ^b$.
Further, we assume that $(\partial_\tau, \partial_\varphi)$ are Killing vectors. 
An illustrative example, using the conformal K\"ahlerian  Pleba\'nski-Demia\'nski  (PD) class, is given by ($x=\cos\theta$)
\begin{eqnarray}
g_{\tau\tau}=\frac{\Delta_r-a^2\Delta_x}{\Pi\Sigma},  g_{\tau\varphi}=a\Bigl[\frac{(r^2+a^2)\Delta_x-(1-x^2)\Delta_r}{\Pi\Sigma}\Bigr],\cr
g_{\varphi\varphi}=\Bigl[\frac{a^2(1-x^2)\Delta_r-(r^2+a^2)\Delta_x}{\Pi\Sigma}\Bigr], 
g_{xx}=-\frac{\Sigma}{\Pi\Delta_x}, g_{yy}=-\frac{\Sigma}{\Pi\Delta_y},\label{3.9}
\end{eqnarray}
with $(\Delta_x, \Delta_y,\Pi(r,x))$ the PD metric functions and $\Sigma =r^2+a^2x^2$\cite{ple1977}. They contain several parameters, such as the mass M, angular  momentum  J and Nut charge N. The Schwarzschild and Kerr solutions are  special cases.
If one introduces complex coordinates
\begin{eqnarray}
z^0=\tau-(r^*-iax^*), \quad z^1=\varphi-(ar^{\#}-ix^{\#}),\quad \quad\cr  dr^*=\frac{r^2+a^2}{\Delta_r}dr,\quad  dx^*=\frac{1-x^2}{\Delta_x}dx, \qquad dr^{\#}=\frac{dr}{\Delta_r},\quad  dx^{\#}=\frac{dx}{\Delta_x},\qquad\qquad\label{3.10}
\end{eqnarray}
the manifold $K$ is conformally K\"ahler with conformal factor $\Pi/(r-iax)^2$.
The K\"ahler form $\hat K$ is then
\begin{eqnarray}
\hat K=\frac{i}{(r-iax)^2}\Bigl[-d\varphi\wedge\Bigl(a(1-x^2)dr-i(r^2+a^2)dx\Bigr)+d\tau\wedge(dr-iadx)  \Bigr],\qquad\label{3.11}
\end{eqnarray}
which is independent of $(\Delta_x,\Delta_y)$. The K\"ahler potential is
\begin{equation}
\upkappa =\int\frac{r}{\Delta_r}dr-\int\frac{x}{\Delta_x}dx.\label{3.12}
\end{equation}
If one takes  the Euclidean  metric 
\begin{equation}
g=\frac{a^2x^2}{4((1+x^2)^2}\begin{pmatrix}
1&\cos y&0&0\\
\cos y &1+x^2\sin^2 y&0&0\\
0&0&1&0\\
0&0&0&1+x^2
\end{pmatrix},
\end{equation}
one proves that the geometry is conformal K\"ahler with 
\begin{equation}
\Omega^2_\pm=\Bigl(\frac{1+x^2}{x^2}\Bigr)^{1\pm\frac{1}{a}}.\label{3.13}
\end{equation}
It is remarkable that this metric is invariant under $(r,M)\leftrightarrow \pm(iax,iN)$. 
The final conclusion is that the FS solution is a gravitational instanton.
\subsection{The Newman-Janis shift}\label{3-2}
Newman et al. \cite{janis1965} discovered a remarkable connection between the Schwarzschild and Kerr solutions. For the special case $x=\cos\theta$, the Schwarzschild solution in Eddington--Finkelstein coordinates is transformed into the Kerr solution by the complex coordinate transformation
\begin{equation}
r'=r+ia\cos\theta,\qquad u'=u-ia\cos\theta.\label{3.15}
\end{equation}
The corresponding Weyl curvature spinors are related as $\Psi_{\mathrm{Kerr}}(x)=\Psi_{\mathrm{Schw}}(x+ia)$,
while their Kähler potentials are connected by the transformation given in Eq.~(\ref{3.15}).
It is remarkable that intrinsic spin can emerge from the Schwarzschild solution through this complexification. A vast literature has been devoted to this subject. For a recent overview, see the review by Kim\cite{kim2025}. We summarize several of the main conclusions.
For the charged Kerr solution of the Einstein--Maxwell equations, the magnetic moment and intrinsic spin of a particle can be interpreted as the projection, or 'shadow', of a particle moving in a complex spacetime onto the physical spacetime. The Dirac gyromagnetic ratio arises when the particle's complex center of charge coincides with its complex center of mass. Consequently, a sufficiently massive charged spinning particle naturally possesses the Dirac gyromagnetic ratio.
There is also a striking analogy between the Kerr solution, with its singular rotating ring of mass, and the electromagnetic field generated by a rotating ring of charge, analogous to the well-known correspondence between the Schwarzschild gravitational field and the Coulomb electric field.
Furthermore, intriguing connections have been established with the classical double-copy construction, in which the Kerr solution is related to the gauge potential of the SD dyon in electromagnetism, the Dirac string, and the peculiar properties of the semi-infinite line mass \cite{kim2025,newman2002,bell2002}. In this framework, the Kerr solution can be interpreted as a pair of SD and ASD Taub-NUT instantons.
In contrast, we shall show that our model can be transformed into a complex manifold in a much simpler manner, without invoking the generation of angular momentum. In our approach, angular momentum is an attribute observed only by an external observer. An observer inside the manifold is free to choose any convenient coordinate system because the angular momentum equation is decoupled from the dynamics of the remaining metric component $N$. Consequently, one may perform the coordinate transformation
$d\varphi \rightarrow d\varphi + iN^\varphi d\tau.$ 
\subsection{The Janis-Newman-Winicour (NJW)  solution and the origin of spin}\label{3-3}
\begin{figure}[h]
	\centerline{
	\includegraphics[width=2.6cm]{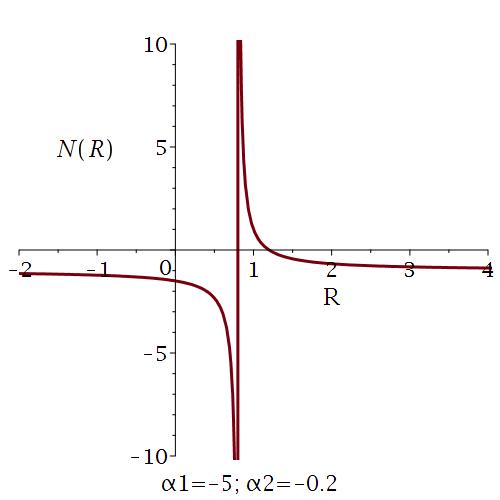}
	\includegraphics[width=2.6cm]{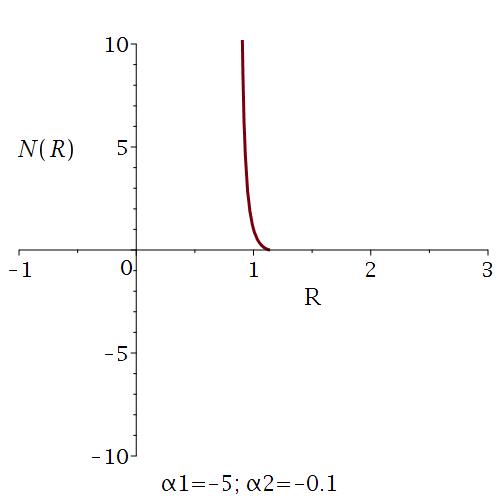}
	\includegraphics[width=2.6cm]{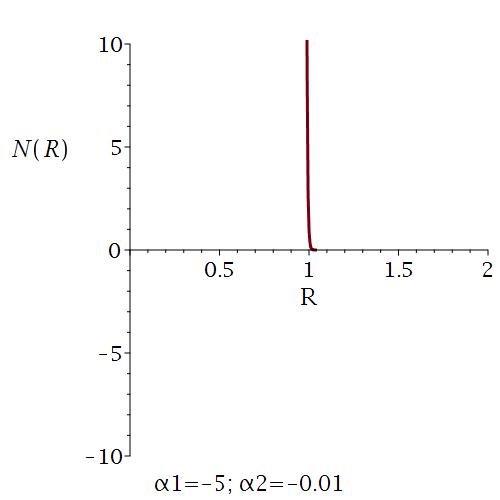}}
	\centerline{
	\includegraphics[width=2.6cm]{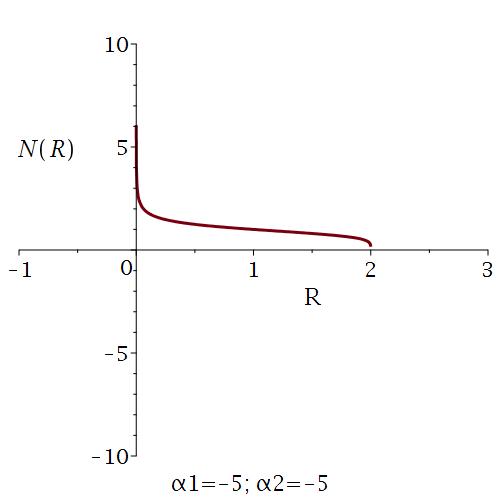}	
	\includegraphics[width=2.6cm]{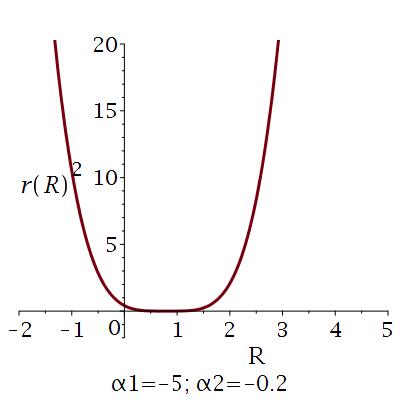}
	\includegraphics[width=2.6cm]{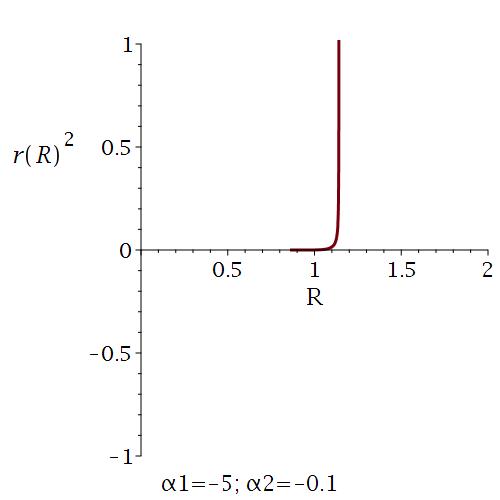}}
	\centerline{
	\includegraphics[width=2.36cm]{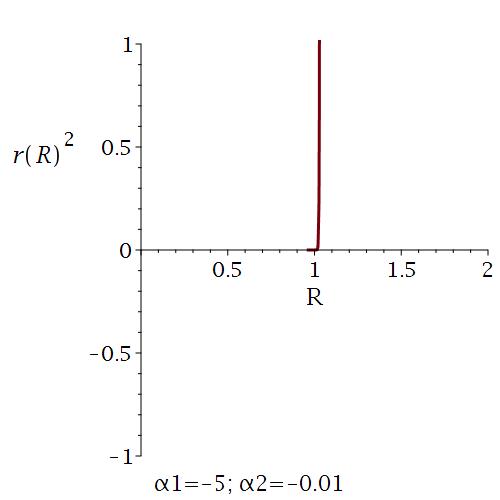}
	\includegraphics[width=2.6cm]{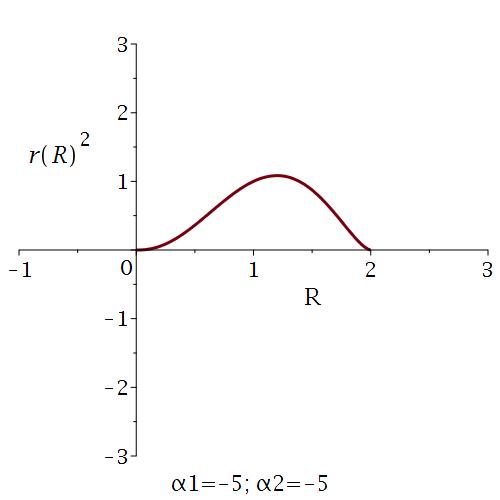}}		
	\caption{{\it Plot of the JNW solution for $M=0.5, \alpha_4 =0, \kappa =1$ and various values of $\alpha_1$ and $\alpha_2$.}} 
	\label{Janis}
\end{figure}

Janis et al., found in their epic study in 1968, an interesting link between the central singularity of the Schwarzschild solution and  the solution with a massless scalar field $\phi$\cite{janis1968}. 
Consider the spacetime
\begin{equation}
ds^2_{JNW}=-N(R)dt^2+\frac{1}{N(R)}dR^2+r(R)^2\Bigl(d\theta^2+\sin^2\theta d\varphi^2\Bigr),\label{3.16}
\end{equation}
with field equations
\begin{equation}
G_{\mu\nu}=-\kappa T_{\mu\nu},\qquad T_{\mu\nu}=\partial_\mu\phi\partial_\nu\phi,\qquad \nabla^2\phi=0.\label{3.17}
\end{equation}
The solution is easily found by GRTENSOR,
\begin{eqnarray}
N(R)=e^{\alpha_4}\Bigl(\frac{\sqrt{\alpha_1\alpha_2}+\alpha_2(R-2M)}{\sqrt{\alpha_1\alpha_2}-\alpha_2(R-2M)}\Bigr)^{1/\sqrt{\alpha_1\alpha_2}},
\quad r(R)=\frac{\alpha_2(R-2M)^2-\alpha_1}{\alpha_2 K(R)},\qquad\cr
 \phi(R)=\sqrt{\frac{1-\alpha_1\alpha_2}{2\kappa\alpha_1\alpha_2}}\ln\Bigl[\frac{\sqrt{\alpha_1\alpha_2}-\alpha_2(R-2M)}{\sqrt{\alpha_1\alpha_2}+\alpha_2(R-2M)}\Bigr],\qquad\qquad\qquad\label{3.18}
\end{eqnarray}
with $\alpha_i$ and M constants. For $\alpha_1\alpha_2 >1$,  $\phi$ becomes complex.
In Fig.(\ref{Janis}) we plotted some characteristics of the solution.
It turns out that the solutions for $(N(R),r(R),\phi(R))$ are completely determined by the Einstein equations; the scalar field equation is therefore redundant, just as in the case of our dilaton field.
A remarkable feature emerges. The surface area of a sphere of constant $R$ scales as $r^2$. When $R=\sqrt{\alpha_1/\alpha_2}+2M$, corresponding to $r=0$, one encounters a singular point. However, the sphere at $R=2M$ is no longer a curvature singularity but instead degenerates into a single point. This indicates a form of asymmetric collapse, similar to that found in models of spinning line sources \cite{slagter2025b}. 
In these models an additional parameter naturally emerges, reflecting the underlying topology. A well-known example is the $\gamma$-metric, which includes both the Schwarzschild solution and the Curzon solution (a horizonless, asymptotically flat spacetime). This family exhibits unusual far-field behaviour: the source is located at $R=0$, possesses higher multipole moments, and displays a directional singularity. In particular, the Kretschmann scalar depends on the direction from which the singularity is approached \cite{griff2006,slagter2026b}. The singularity may be interpreted as a ring on which certain timelike geodesics terminate.
Related examples, such as the C-metric and the infinite line-mass solution, demonstrate that, as the mass parameter increases, the integration constant appearing in the solutions yields flat spacetime for two distinct parameter values. As noted earlier, asymptotically flat static solutions with a point singularity, finite mass, and a small degree of asymmetry approach this truncated Schwarzschild solution in the limit where the asymmetry vanishes.
Another illuminating framework is provided by the Kerr Schild (KS) class of metrics \cite{kim2025}, in which the metric is written as
$g^{\mu\nu}=\eta^{\mu\nu}+\lambda^2 l^\mu l^\nu$,
where $l^\mu$ is a null geodesic vector field \cite{flah1976}. This formalism makes the analogy with electrodynamics particularly transparent. One effectively replaces the Schwarzschild radius,
$r_S=\frac{\kappa M}{4\pi},$ by the electromagnetic quantity $gq/4\pi$. In electrodynamics, the gauge field is generated by a prescribed current distribution. Similarly, a KS metric may be viewed as arising from an effective source: for the Kerr solution, this is a rotating disk whose mass distribution gives rise to the ring singularity, whereas for the Schwarzschild solution it corresponds to the NJW phantom source. It should be recalled that the FS solution requires the presence of a cosmological constant.
Furthermore, the Kerr solution can be obtained by complexifying the Schwarzschild geometry and performing a complex coordinate shift, thereby introducing angular momentum. In an analogous manner, a minimally coupled scalar field may be transformed into a spinning particle through an appropriate complexification procedure.
The 3D Kerr-like spacetime is
\begin{equation}
ds^2=[d(t+J\varphi)]^2+[dr^2+r^2d\varphi^2],\label{3.21}
\end{equation}
with $0\leq\varphi\leq2\pi(1-M), T^{00}\sim M\delta^2(r), T^{0i}\sim J\epsilon ^{ij}\partial_j\delta^2(r)$.
It is manifestly locally flat outside the origin, where a singularity represents a massive spinning point source. The connection with line-like structures then becomes apparent, particularly after the transition to complex coordinates, which naturally relates orbital angular momentum to intrinsic spin.
Furthermore, there is a close correspondence with the Dirac string and its associated line singularity. The magnetic dipole moment, with gyromagnetic ratio $e/mc$, is mapped to a complex center of charge line. Likewise, the complex center of mass lines of the Schwarzschild and Kerr geometries differ only by an imaginary displacement. As emphasized by Newman, both a momentum vector and an angular momentum tensor can be defined in $\mathcal{H}$-space ('Heaven') from the asymptotic curvature data at null infinity of the original manifold. The imaginary displacement of the center-of-mass line signals the presence of intrinsic spin, so that the angular momentum of the Kerr solution is naturally interpreted as intrinsic spin in the complexified geometry. In this sense, the Kerr spacetime exhibits many of the characteristics of a spinning particle\cite{hamed2019}.
As we shall demonstrate, our solution is closely related to this class of geometries. In contrast to the original JNW model, which is commonly interpreted as describing a wormhole-like configuration, our construction does not require such an interpretation. Consequently, the problem of the large mutual gravitational attraction between two black holes is avoided. Instead, the geometry is supported by a novel global topology, allowing it to be naturally connected with (anti-) self-duality and the structure of a conformal K\"ahler manifold.
\section{The new instanton}\label{sec:4}
The concentration of instantons in an annular region of $S^4$ can be interpreted as a conformal deformation in the five-dimensional setting. Through Hawking radiation these localized configurations may gradually decay and eventually evaporate.
Following the theorem of Taubes, we assume that SD connections on the four-manifold $M$ arise from SD solutions on $S^4$, whose moduli space is denoted by $M_5$. In our construction, the topology of the compactified moduli space is identified with that of the five-dimensional warped spacetime.
A YM instanton on $S^4$ is characterized by its center $b\in S^4$ and its scale parameter $\Omega\in\mathbb{R}^+$. As $\Omega\rightarrow 0$, the instanton becomes increasingly localized near $b$. Taubes construction grafts this localized SD connection onto $M$, where the resulting connection initially acquires only a small ASD component. For sufficiently small $\Omega$, this error can be removed by a perturbation, yielding an exact SD connection. These solutions form a collar neighborhood of $M$ inside the compactified moduli space $M_5$.
Our approach differs from Taubes construction because the action is conformally invariant. Consequently, the instanton scale is naturally replaced by the conformal factor, while the elliptic field equations remain identical on both $M$ and $M_5$.
Consider an instanton represented by a point in the moduli ball $B^5$. Let
$T_{\Omega,b}:x\mapsto \Omega(x-b)$ denote a local conformal transformation. We identify antipodal configurations through the equivalence
$T_{\Omega,b}\sim T_{1/\Omega,b^*}$, where $b^*$ is the antipodal point of $b$. The transformed instanton $T_{\Omega,b}^*A$ therefore has center $b$, conformal scale $\Omega$, and a spatial extent determined by $\Omega$. As $\Omega\rightarrow 0$, the curvature approaches a delta-like concentration at $b$.
In the compactified moduli space, points of the boundary $S^4=\partial B^5$ therefore correspond to ideal SD connections whose curvature is represented by a $\delta$-distribution. The compactification of the moduli space by these ideal instantons follows the construction of Freed and Uhlenbeck \cite{freed1984}.
\subsection{The new topology}\label{4-1}
\begin{figure}[h]
\centerline{	
	\includegraphics[width=4.cm]{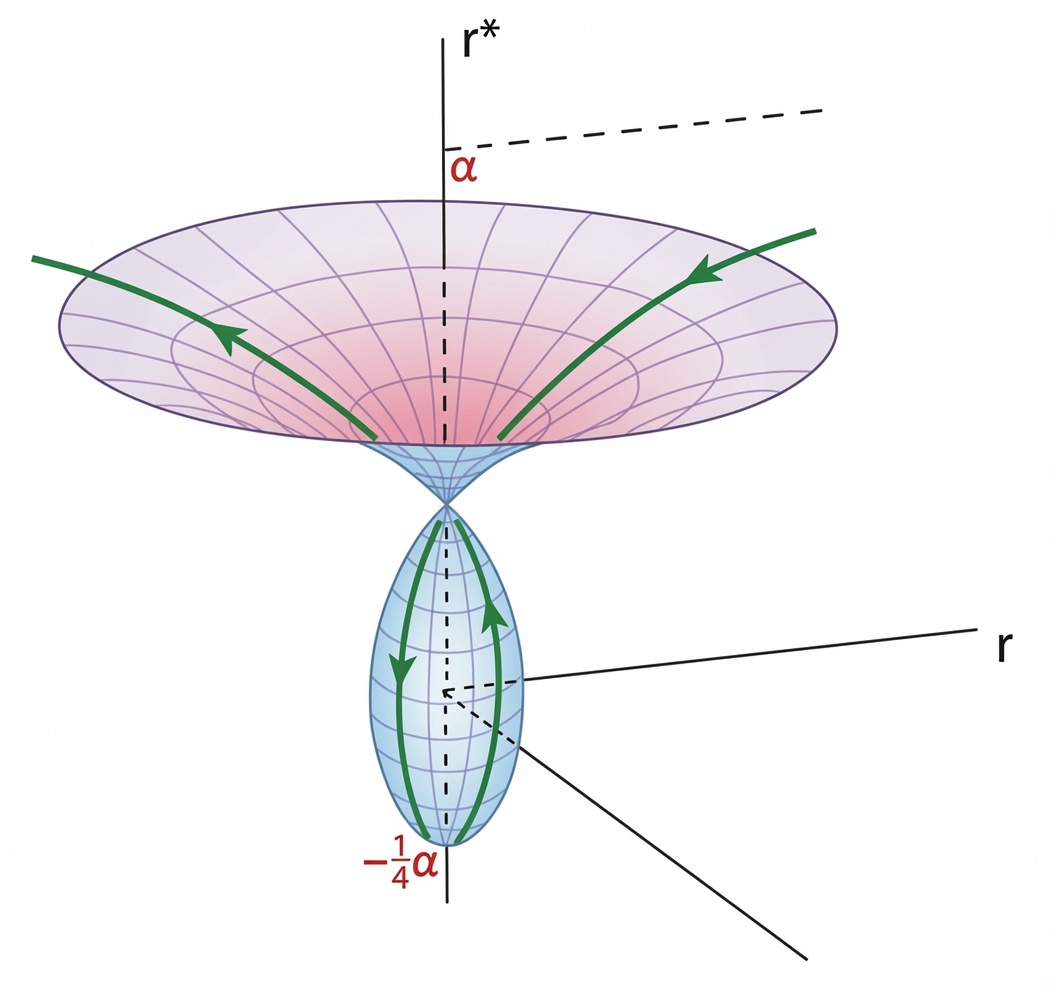}
	\includegraphics[width=5.cm]{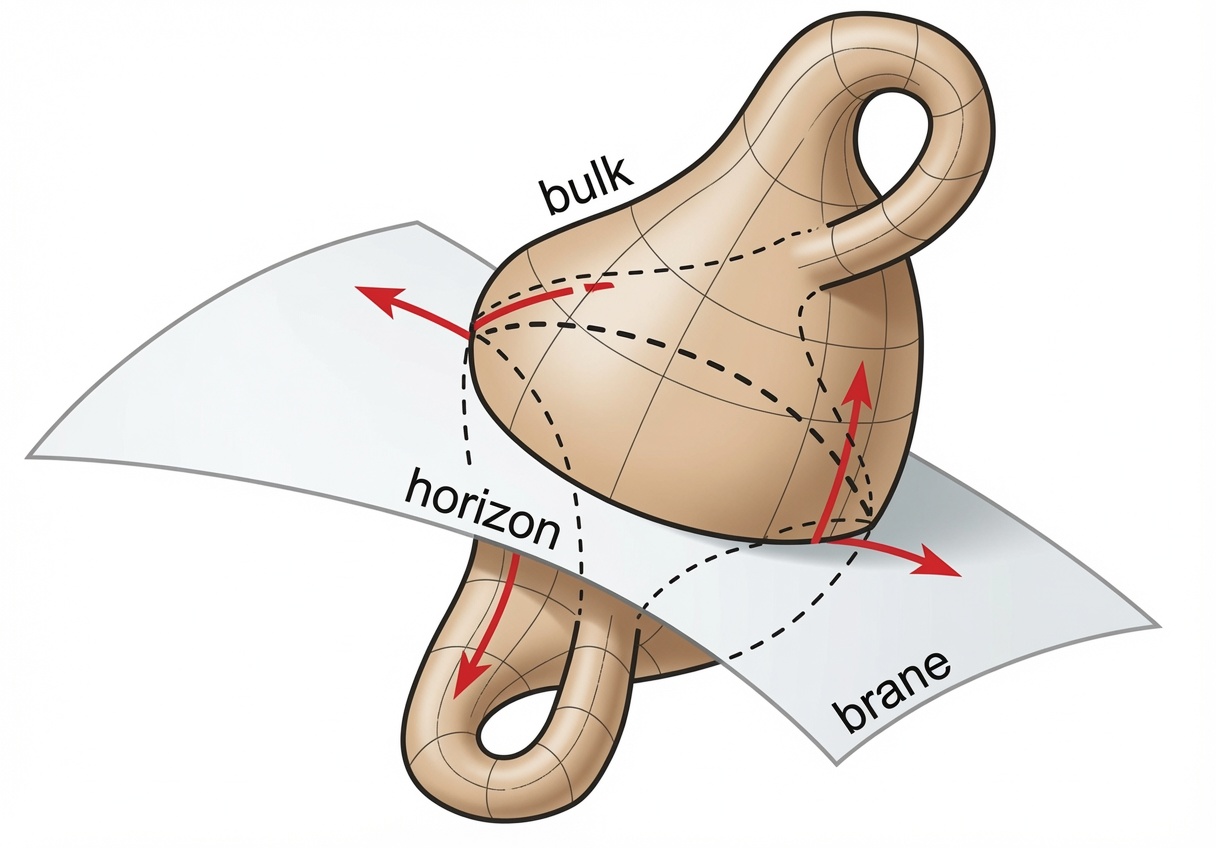}}
	\centerline{
	\includegraphics[width=5.6cm]{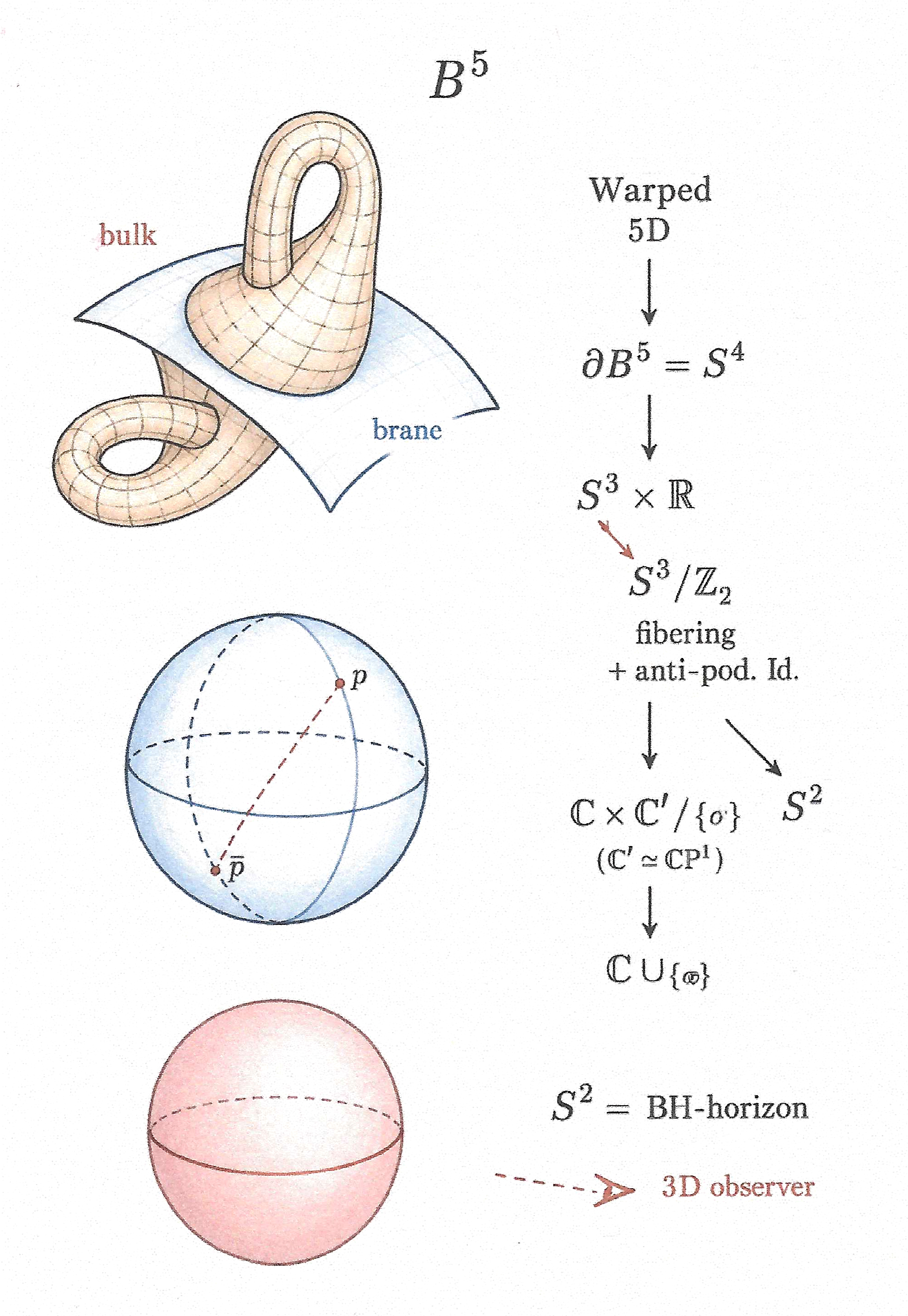}}
	\caption{Top left: Surface of revolution of the interior. Top right:The Klein bottle with $\mathbb{Z}_2$ symmetry. Bottom: The topology of the new black hole. }		
	\label{fig4}
\end{figure}
Let us now consider our  Riemannian stationary effective metric for the special case $C_1=0$,
\begin{equation}
ds_{eff}^2=N(R)^2\Bigl[d\tau^2+dR^2\Bigr]+dz^2+r(R)^2d\varphi^{*2},\label{4.1}
\end{equation}
with 
\begin{equation}
R=\frac{4}{125a^2}\ln\Bigl[\frac{(4r+a)}{(r-a)}\Bigr]+\frac{1}{75a}\frac{17a^2-39ar-3r^2}{(r-a)^3},\label{4.2}
\end{equation}
or
\begin{eqnarray}
ds^2_{eff}=\frac{(a+4r)(r-a)^4}{5r^2}d\tau^2+dr^{*2}+dz^2+r^2d\varphi^{*2},\label{4.3}
\end{eqnarray}
with 
\begin{equation}
r^*=\pm\frac{1}{5\sqrt{a}}\Bigl[ 6 arctanh\Bigl(\sqrt{\frac{a+4r}{5a}}\Bigr)
+\frac{\sqrt{5a(a+4r)}}{r-a}\Bigr],\label{4.4}
\end{equation}
and with $-\frac{a}{4}<r<a$. It  describes the interior of the black hole (Fig.(\ref{fig4})).
We employ a two-fold Riemannian covering by $\mathbb{C}P^2$ together with an antipodal identification on the hypersurface of a Klein bottle, locally represented as $\mathbb{C}^1\times\mathbb{C}^1$, to model the Hawking quanta produced during black-hole evaporation. Through the Hopf fibration, the event horizon is identified with the base manifold $S^2$, while the corresponding gravitational instanton is assumed to possess the topology $S^3\times\mathbb{R}/\mathbb{Z}_2$.
We further conjecture that the metric on the Klein bottle $K^2$, viewed as part of the product manifold $H^2\times S^1$, is SD and K\"ahler, where $H^2$ denotes a compact two-dimensional manifold with boundary $S^1$. This assumption provides a natural geometric setting for the conformal description of the instanton.
It is well known that every complex manifold admitting a holomorphic embedding into complex projective space carries a K\"ahler metric. Moreover, if $(M,g)$ is a Hermitian manifold with an integrable almost-complex structure, one may ask whether there exists a conformal factor $\Omega$ such that the conformally related metric $\hat g=\Omega^2 g$
is K\"ahler. Such manifolds are known as globally conformally K\"ahler when a global conformal factor exists, or locally conformally K\"ahler when the conformal K\"ahler structure exists only locally.
\subsection{Conformally K\"ahler form}\label{4-2}
The first block of our metric describes the axisymmetrix space
\begin{eqnarray}
ds_{2D}^2=N(r)^2 d\tau^2+\frac{1}{N(r)^2}dr^2= \partial_\xi\partial_{\bar\xi}\upkappa d\xi \wedge d\bar\xi,\label{4.5}
\end{eqnarray}
with $\xi=\rho(r)e^{i\tau}$ (the 'disk' method) and (see Eq. (\ref{4.1}))
\begin{eqnarray}
C(r)\equiv\log|\rho(r)|=\int \frac{1}{N^2} dr=R.\label{4.6}
\end{eqnarray}
$\upkappa =J(\rho)=J(\sqrt{\xi\bar\xi})$ is the K\"ahler potential. One finds finally\cite{fre2012}
\begin{equation}
\frac{dC}{dr}=\frac{1}{N^2}\qquad \frac{d^2 J}{dC^2}=4N(C)^2.\label{4.7}
\end{equation}
So one needs the inverse $r=r(C)$.
The K\"ahler metric is $ds_2^2=\partial_\xi\partial_{\bar \xi}\upkappa d\xi d\bar\xi$.
Another possibility is to write (the 'plane' case) (see Eq. (\ref{4.3}))
\begin{equation}
ds_2^2=\frac{(a+4r)(r-a)^4}{5r^2}d\tau^2+dr^{*2}.\label{4.8}
\end{equation}
One replaces $\xi\mapsto \xi(\rho(r^*),\tau)\equiv iC(U)-\tau$. The K\"ahler potential is now a function, only of the imaginary part of $\xi$, i.e.,
$\upkappa(\xi,\bar\xi)=J(Im\xi)=J(C)$. The explicit form of the 2D metric becomes
\begin{equation}
ds_2^2=\frac{1}{4}\frac{\partial^2J(C)}{\partial C^2}\Bigl(\frac{\partial C(U)}{\partial C}\Bigr)^2dC^2+\frac{1}{4} \frac{\partial^2J(C)}{\partial C^2}d\tau^2.\label{4.8b}
\end{equation}
In this case, the coordinate $\tau$ is compact on $(0,2\pi)$, while in the plane case over $\mathbb{R}$. 
In our model this is equivalent to the change of the $z$ coordinate to the bulk coordinate, which becomes an angle variable by the Klein bottle topology.
The equation for $J$ becomes
\begin{equation}
\frac{d^2J}{dC^2}=4N(r^*)^2, \qquad \Bigl(\frac{dC}{dr^*}\Bigr)^2\frac{d^2J}{dC^2}=4.\label{4.9}
\end{equation}
So $\quad \frac{dC}{dr^*}=\frac{1}{N(r^*)}$.
Further, one can evaluate 
\begin{equation}
\frac{dJ}{dC}=\int N(r^*)^2 dC =\int N(r^*)^2\frac{dC}{d r^*}dr^* =
\int N(r^*) dr^*.\label{4.10}
\end{equation}
The K\"ahler metric possesses an isometry for Killing vectors $k^\zeta$,
\begin{eqnarray}
\zeta\rightarrow \zeta + k^\zeta,\quad \bar\zeta \rightarrow \bar\zeta +k^{\bar\zeta}, k^\zeta(\zeta)=ig^{\zeta\bar\zeta}\partial_{\bar\zeta}P,  \quad  k^{\bar\zeta}(\bar\zeta)=-ig^{\bar\zeta\zeta}\partial_{\zeta}P.\label{4.11}
\end{eqnarray}
For some real $P(\zeta,\bar\zeta)$ we then have
\begin{equation}
P=-\frac{1}{2} i\Bigl(k^\zeta\partial_\zeta \upkappa -k^{\bar\zeta}\partial_{\bar\zeta}\upkappa \Bigr).\label{4.12}
\end{equation}
The K\"ahler potential is also  invariant under the isometry.
There is also an isometry of $\tau \rightarrow \tau +a$, generated by $k^\zeta=i\zeta, k^{\bar\zeta}=-i\bar\zeta$. We then have
\begin{equation}
P= \rho\frac{dJ}{d\rho}=\frac{dJ}{dC}.\label{4.13}
\end{equation}
In the 'plane' case, we have for a translation, $k^\xi=1$. So again
\begin{equation}
P=-\frac{1}{2} i\Bigl(k^\xi\partial_\xi \upkappa -k^{\bar\xi}\partial_{\bar\xi}\upkappa \Bigr)= \frac{dJ}{dC}.\label{4.14}
\end{equation}
So in this 'plane' case, we conclude that 
\begin{equation}
N(r^*)=\frac{d}{dr^*} P(r^*).\label{4.15}
\end{equation}
Our $N^2$  is the derivative  with respect to $r^*$ of the 'momentum map' of the Killing vector which generates translations of the cyclic $\tau$.
Next, we include the second block.
An odd-dimensional manifold, however, does not admit a Kähler form. Since the effective manifold possesses the same solutions, apart from the dilaton solution, we may instead work with the corresponding four-dimensional effective manifold.
It has been shown that type D vacuum and Einstein-Maxwell spacetimes admit a Hermitian structure in the Riemannian setting and satisfy the Lorentzian analogue of the conformal K\"ahler condition\cite{flah1976}. Motivated by this result, we write our four-dimensional effective Euclidean manifold as
\begin{equation}
ds_{eff}^2=N(r)^2d\tau^2+\frac{1}{N(r)^2}dr^2+dz^2+r^2d\varphi^{*2},\label{4.16}
\end{equation}
with $d\varphi^*=d\varphi+iN^\varphi(r)d\tau$, in block form as ($t\rightarrow i\tau, N^\varphi\rightarrow iN^\varphi$)
\begin{equation}
ds^2=a_{ij}d\sigma^i d\sigma^j +b_{IJ}dx^I dx^J,\label{4.17}
\end{equation}
with $\sigma^i=(\tau,\varphi), x^I=(r,z)$.
We also want to write our manifold in conformal K\"ahler form
\begin{equation}
\bar g_{i\tilde j}=\Omega_k^2 g_{i\tilde j},\qquad  \bar g_{i\tilde j}=\partial_{z^i}\partial_{\tilde z^j}\upkappa,\label{4.18}
\end{equation}
where $\upkappa$ is the K\"ahler potential, which was constructed in the preceding section for the first block of the metric and $\Omega_k$ the conformal factor (not to confuse with $\Omega$ from section 2). 
In general we have $p_j=\partial\upkappa/\partial \tilde z^j=\int\bar g_{i\tilde j}dz^i$. Integrating once again, we obtain $\upkappa=\int p_i d\tilde z^i$.
We observed that $N^2(r)$ is twice integrable, with 
\begin{equation}
\upkappa=r^2\Bigl(\frac{1}{25}r^3-\frac{a}{4}r^2+\frac{2a^2}{3}r-a^3\Bigr)-\Bigl(\frac{a^5}{5}+C_1\Bigr)\ln r +\alpha_1 r+\alpha_2,\label{4.19}
\end{equation}
with $\alpha_i$ integration constants.  This function is every where regular for suitable values of the constants. 
One now proves that the conformal K\"ahler condition is\cite{aksteiner2022} $d(\Omega^2 K)=0$, with $K$ the 2-form.
Our total effective 4D manifold (note that we have still the conformal invariance factor $\Omega^2$) becomes
\begin{eqnarray}
ds^2_{eff}=\omega^2\Bigl[N(r)^2d\tau^2+\frac{1}{N(r)^2}dr^2+dz^2+r^2d\varphi^{*2}\Bigr]\cr =\omega^2\Bigr[
N(R)^2\Bigl(d\tau^2+dR^2\Bigr)+dz^2+r(R)^2d\varphi^{*2}\Bigr].\label{4.20}
\end{eqnarray}
Next, we write the second block $(dz, d\varphi^*)$
\begin{eqnarray}
dz^2+r^2d\varphi^{*2}=r^2(\frac{1}{r^2}(dz^2+d\varphi^{*2})\qquad\cr=r^2\Bigl(\frac{tan^2\theta}{z^2}dz^2+d\varphi^{*2}\Bigr)=r^2\Bigl(tan^2\theta dz^{*2}+d\varphi^{*2}\Bigr),\label{4.21}
\end{eqnarray}
where we isolated a conformal factor $r^2$. Further, $z\equiv e^{-z^*}$ and $\theta$ constant, because we work in the plane.
Finally we arrive at our 2-form, by adding the first block of Eq.(\ref{4.5})
\begin{equation}
\tilde K=\Omega(r)^2 K=
\Omega(r)^2\Big[\partial_\xi\partial_{\bar\xi}\upkappa_1d\xi\wedge d\bar\xi  +\partial_\chi\partial_{\bar\chi}\upkappa_2d\chi \wedge d\bar\chi\Bigr].\label{4.22}
\end{equation}
We isolated the $r^2$-term and absorbed it in the conformal factor.
Further, by suitable choice of $\Omega$, we can obtain $d(\Omega^2 K)=0$. 
Remember that we still have the dilaton factor $\sim\omega^2$ in front of $\bar g_{\mu\nu}$. 
We have proven that our solution is a gravitational instanton, comparable with the FS.
We can write our effective 4D manifold in compact two-form
\begin{eqnarray}
K=\partial_\zeta\partial_{\bar\zeta}\upkappa d\zeta d\bar\zeta,\qquad \zeta=\{\xi(\tau,r),\chi(z,\varphi)\}, \quad \xi=\rho(r) e^{i\tau},\quad \chi=z+i\varphi^*,\qquad \label{4.23}
\end{eqnarray}
with  $\upkappa$ the K\"ahler potential. 
We conclude with a remark concerning the integration constant k appearing in the solution, Eq.(\ref{2.9}). In view of Pleba\'nski's classification, k appears to be quantized and therefore takes integer values. This interpretation is further supported by the analysis of Calabi \cite{calabi1979}, who demonstrated that, in the case of the FS metric, the first- and second-order differential equations are mutually consistent-what he referred to as 'une heureuse coïncidence'. The same remarkable consistency is found in our model, providing additional evidence for the validity of the construction.
The theorem of Calabi on the construction of Ricci-flat K\"ahler metrics on Calabi-Yau manifolds imposes strong constraints on both the topology and the complex geometry of the underlying manifold. In general, the resulting solutions contain two integration constants, one associated with the mass parameter and the other with the topology. These constants arise from the nonlinear elliptic partial differential equation obtained from the Einstein equations. The proof of Calabi's ansatz is highly nontrivial; the existence theorem was ultimately established by Yau. For a comprehensive discussion, we refer to Joyce \cite{joyce2007}.
One may also apply Pleba\'nski's method to our effective four-dimensional Einstein equations. For an overview of this formalism, we refer to Krasnov \cite{krasnov2020}.
For a suitable tetrad $(e^\tau, e^r, e^z, e^\varphi)$, one constructs the SD and ASD two-forms and three-forms,
\begin{eqnarray}
\Sigma^i/\bar\Sigma^i=i e^0\wedge e^i\mp\epsilon^i_{jk} e^j\wedge e^k, \qquad
d^A\Sigma^i\equiv \Sigma^i+\epsilon^i_{jk}A^j\Sigma^k=0,\label{4.24}
\end{eqnarray}
where the $\Sigma^i$ satisfy the self-duality condition. The connection one-forms $A^i$ are obtained from the SD part of the torsion-free spin connection. Once the $A^i$ have been determined, the curvature two-forms
$F^i=dA^i+\frac12\epsilon^i_{jk}A^j\wedge A^k$ can be calculated. They are SD and can therefore be expanded as $F^i=M^{ij}\Sigma^j$.
By requiring that the sum of the diagonal SD components vanishes, one immediately recovers, for example in the Schwarzschild case, the corresponding first-order differential equation for the metric function. For the FS instanton one finds
$F^i=-\frac{\Lambda}{3}\Sigma^i,$ while the Weyl curvature satisfies $\Psi^{11}=0$, thereby providing a direct proof that the FS solution is a gravitational instanton.
In our case, we  write
\begin{equation}
G_{\mu\nu}-{\cal E}_{\mu\nu}=\frac{1}{\omega^2}\Bigl[T^{(\omega)}_{\mu\nu}-\lambda\kappa^2\omega^4 g_{\mu\nu}\Bigr].\label{4.25}
\end{equation}
Note that the left-hand side contains the trace-free tensor ${\cal E}_{\mu\nu}$, which is an essential ingredient of the Pleba\'nski formulation. Accordingly, we also decompose the dilaton energy-momentum tensor into its trace-free part, 
\begin{eqnarray}
\bar T^{(\omega)}_{\mu\nu}=T^{(\omega)}_{\mu\nu}-\frac{1}{4}g_{\mu\nu}T^{(\omega)},\qquad T^{(\omega)}=\frac{6}{ry_0}\Bigl([rN^2\omega\omega']'-rN^2[\omega']^2\Bigr), \label{4.26}
\end{eqnarray}
so that $\bar T^{(\omega)}_{\mu\nu}$ is manifestly trace-free. 
The Pleba\'nski formalism therefore provides a particularly elegant framework for establishing the self-duality of our solution \cite{slagter2026b}.
\subsection{Relation between the quintic and Calabi-Yau manifold}\label{4-3}
Our quintic can be transformed into several well-known canonical forms. For a special value of the discriminant, it assumes the Tschirnhaus form
$z^5+\frac{15}{16}a(c-a)^5r^2+\frac{125}{256}a^3(c-a^5)z-\frac{1}{16}(c-a^5)^2$,
the canonical form
$z^5+5az^2+5bz+c$,
the Brioschi form
$z^5-10\beta z^3+45\beta^2z-\beta^2$,
and the Bring-Jerrard form
$z^5+az+a$\cite{slagter2022}. These representations reveal remarkable connections between the quintic equation, the symmetry group of the icosahedron, elliptic curves, and the Riemann sphere. Moreover, there exists a profound relationship with the modular group and the rational curves of Calabi-Yau manifolds \cite{can1991}.
Expressing the quintic in its factorized form,
$N=\prod_{i=1}^{5}(z-z_i),$ allows one to study the evolution ('dance') of the roots (see Fig.\ref{fig1}) as the parameter (a) becomes small. This behaviour is closely related to the results of \cite{can1991}. In the Brioschi form, the ordered roots correspond to points on the icosahedron. One may then construct three invariant homogeneous polynomials associated with the vertices, faces, and edges, denoted by (V), (F), and (E), respectively. These satisfy the classical icosahedral relation
$1728V^5-F^3-E^2=0$. The solution of the quintic can therefore be formulated in terms of the icosahedral equation and the associated elliptic curves. For a comprehensive treatment of this elegant correspondence, we refer the reader to the monograph by Toth \cite{toth2002}.
\subsection{Our complex transformation}\label{4-4}
\begin{figure}[h]
	\centerline{
	\includegraphics[width=4.cm]{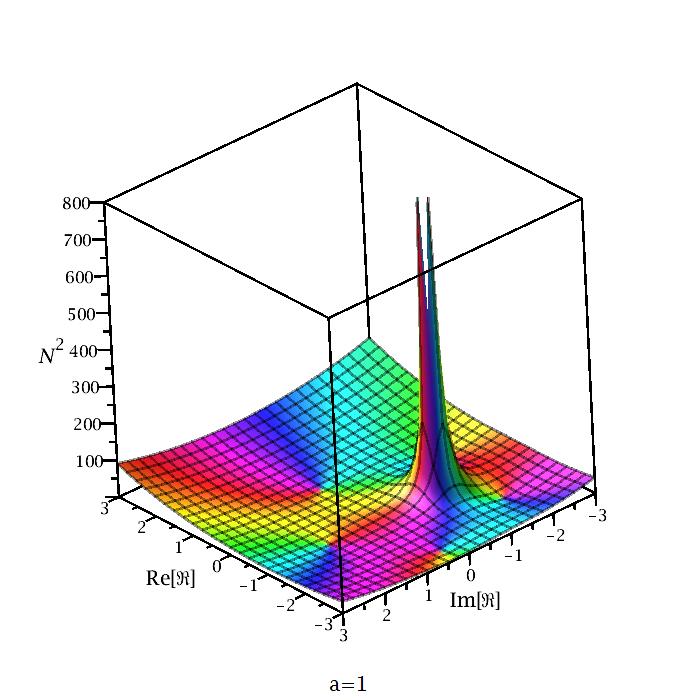}
	\includegraphics[width=4.cm]{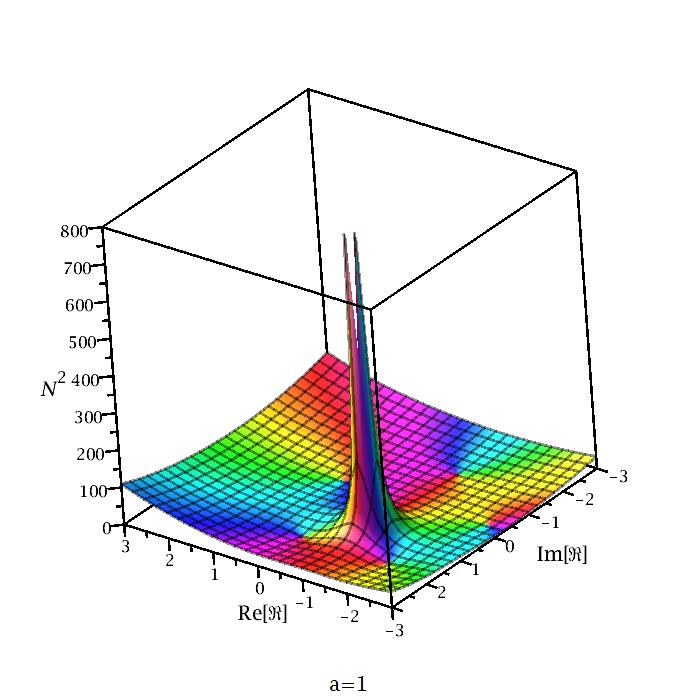}}
	\centerline{
	\includegraphics[width=4.3cm]{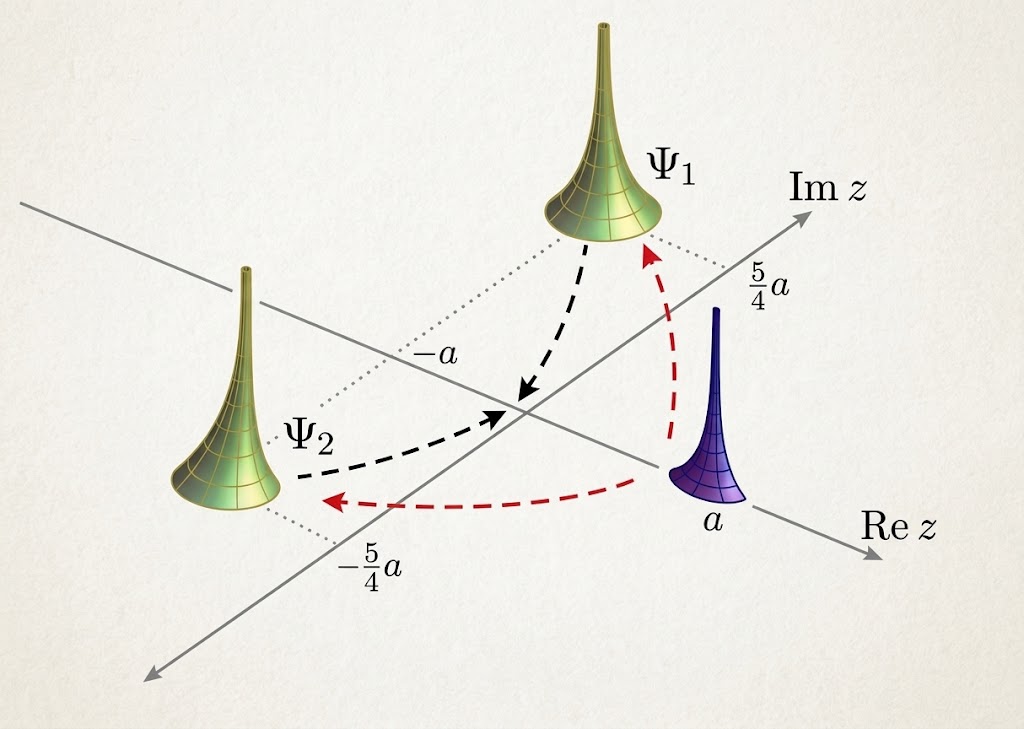}}
\caption{{\it Top: Plot of the  two branches of the complex metric component $N^{\pm}(R)$, for $C=3381 a^5/256$ and $a=1$.
Bottom: When a decreases, the two complex zeros approach $z=-a$, so the central singularity is approached  from the negative z-side.}} 
	\label{fig5}
\end{figure}

Our model can be shifted to the complex plane, in analogy with the EH case, by replacing the singular term in the polynomial, $(r-a)^4)$, with the complexified expression $(R+a)^4$. This corresponds to the transformation $R=-a\pm i(r-a)$.
In Eq. (\ref{4.20}), the corresponding factor becomes complex, and its behavior is regular throughout the complex plane, except at the points
$(-a,-a\pm \frac{5}{4}ia)$. The singularities are located at $R=-a\pm ia$,
as illustrated in Fig.(\ref{fig5}). In the limit $a\rightarrow 0$, the two complex zeros move towards $z\uparrow 0$ from the complex direction.  One could relate $a$ to the 'mass' of the instanton.
The dilaton contribution transforms accordingly, becoming
$\omega^2=-\frac{b^2}{(R+a)^2}$. 
The complexified metric component $N(R)$ then takes the form
\begin{equation}
N^2_{comp}(R)=\frac{(R +a)^4(5a\pm 4i(R+a))}{5(a\pm i(R +a))^2}.\label{4.27}
\end{equation}
In Fig.~(\ref{fig5}) we plotted both branches.
The Kähler potential can also be obtained for the first block from the relation
$\partial_R\partial_{\bar R}\upkappa=N_{\mathrm{comp}}(R)$. In addition, we require a complex shift $\varphi\rightarrow\varphi+iN^\varphi dt$.
This transformation is possible because the equation determining $N^\varphi$ decouples from the remaining metric components.
An other issue is to provide a consistent topological interpretation of the JNW trick. A possible explanation may be obtained from a description in terms of a pair of SD and  ASD Taub-NUT or Kerr--Newman instantons. However, we prefer to retain the simplicity of our vacuum model. In this framework, we obtain an interior solution with only non-essential singularities.
This interpretation is related to the viewpoint that the origin of the complex shift is a consequence of spin effects arising when one passes from a minimally coupled scalar field (the dilaton) to a spinning particle\cite{hamed2019}. In this context, it is natural to seek a connection with string-like structures and with Kerr-type solutions in $(2+1)$-dimensional models. A representative metric is
$ds^2=[d(t+J\varphi)]^2+dr^2+r^2d\varphi^2,\qquad 0\leq\varphi\leq 2\pi(1-M)$.
For a spinning source, the matching conditions are determined by both a deficit angle and a time shift\cite{deser1992}. This suggests that the quantization of angular momentum may be related to discrete identifications or jumps in the time coordinate.
Next, we aim to establish a link between the formation of the two branches and a quantum entangled pair. Could we identify an entanglement measure? Or even a description of the borderline between quantum behaviour on a very small scale and the development of a GRT black hole? Will the pair of instantons remain in a pure superposition state? 
What about hidden variables versus non-locality? Is super-determinism possible? 
A huge amount of literature exists on this part of QM.  Many attempts have been made in situations where the background gravitational field is flat or an approximation method is applied. 
Could we not investigate the possibility of superposition of the metric itself\cite{penrose2014}? Quoting Penrose: {\it 'Nature dislikes and therefore suppresses superposition of locally different spacetime curvatures. The more massive the system in superposition is the greater the difference between the two spacetimes and the faster the collapse of the wave function'}. The expected time is $\Delta \tau\sim \hslash/\Delta E_G$, where $\Delta E_G$ can be seen, by applying Heisenberg, as the uncertainty in the energy of the superposed state.
In our case we have on the right hand side the contribution from the projected Weyl tensor of the bulk 
\begin{eqnarray}
{\cal E}_t^t=\frac{1}{24y_0}\Bigl[4NN''+4(N')^2+4r^2N^{(\varphi)}N^{(\varphi)''} +3r^2(N^{(\varphi)'})^2+12rN^{(\varphi)}N^{(\varphi)'}\Bigr].\label{4.28}
\end{eqnarray}
If we apply the transformation to $\varphi^*$ for the local observer, we have the contribution $(NN')'/6y_0$
\subsection{Super-determinism, wave function collapse and Fubini-Study distance }\label{4-5}
One of the most widely studied controversies is undoubtedly Bell?s theorem. Do hidden variables exist in our world, or must we accept that superluminal transport occurs in quantum mechanics?
We have, globally,  four possibilities. Firstly, reject Bell?s hidden-variables hypothesis. Secondly, locality is violated in our world. Thirdly, The world is super-deterministic and finally there exists super correlation.
Many physicist now believe that super determinism is a promising approach to solve not only the measurement problem but also to understand the apparent non-locality in QM.
Let?s see what we can say about this controversy in our model of a black hole\cite{slagter2025b}. 
We will follow, more or less, 't Hooft?s approach: {\it 'all atoms and photons are in a true ontological state and form an ontological basis. The universe began in such a state and the laws of evolution are such that it will be in this state at any moment in the future. There is no quantum superposition of different ontological states'.}
This sensitive issue is also linked to the concept of free will. Clearly, we will not be discussing this topic in greater depth here. 
In any case, we know this from the two-quasar experiment, in which two photons from quasars more than two billion light-years away are used to determine the orientation of the polarisation detectors just prior to the arrival of entangled photons  at Alice and Bob. Clearly, photons have no free will, so the question of free will is deferred by a couple of billion years.
In our model, we have an exact physical solution, albeit not yet in the quantum realm. The $\omega$ field is close to the horizon and is a 'normal' gravitational component. As the scale becomes smaller, $\omega$ becomes a quantum field.
Our manifold is part of a K\"ahler manifold and resembles a complex projective, conformal, flat FS manifold.  This means that, rather than having a purely flat metric as is usually assumed when calculating the FS distance, we have a complex conformal factor.
One might ask, 'What is the distance between two points in quantum state space?' A geometric interpretation of statistical distance can indeed be defined by calculating the number of distinguishable states\cite{whooter1981}.  In the words of Whootters, {\it 'A concept of statistical distance is defined between different preparations of the same quantum system, or, in other words, between different rays in the same Hilbert space'. Statistical distance is determined entirely by the size of the fluctuations occurring in the measurements designed to distinguish one state from another.' It is not related, a priori, to the usual distance (or angle) between rays.' However, one finds that these two kinds of distance are in fact the same, a result which depends on certain peculiarities of quantum mechanics'}.
If one has 2 quantum states $\ket{\psi_1}$ and $\ket{\psi_2}$ with probability distributions $p_i^{(1)}, p_i^{(2)}$ under some measurements basis, then the distance for distinguishing two probability distributions can be defined by
\begin{equation}
d_{stat}(p^{(1)},p^{(2)})=\arccos\Bigl(\sum_i \sqrt{p_i^{(1)}p_i^{(2)}}\Bigr).\label{4.29}
\end{equation}
One conjectured that the FS metric is suitable to define a metric in Hilbert space and equals to the classical statistical distance under optimal measurement.
The geodesic distance of the two states $\ket{\phi_1}$ and $\ket{\phi_2}$ on a projective space $\mathbb{C}P^n$ is then
\begin{equation}
s(\phi_1,\phi_2)=\arccos|\braket{\phi_1|\phi_2}|.\label{4.30}
\end{equation}
One can also use a given curve in the projective Hilbert space H and try to measure  distances by the FS metric. The next step is then to find a kind of entanglement distance. 
In most model one applies an approximation to obtain a distance between state vectors in H, say, $\ket{\psi_1}$ and $\ket{\psi_2}$. One defines $D(\ket{\psi_1},\ket{\psi_2})=|\ket{\psi_1}-\ket{\psi_2}|=\braket{\psi|\psi}^{1/2}=[2(1-Re(\braket{\psi_1|\psi_2}))]^{1/2}$, with $\ket{\psi}=\ket{\psi_1}-\ket{\psi_2}$. To second order, one could obtain the squared infinitesimal distance between two neighboring vectors, i.e., $d^2(\ket{\psi}+\ket{d\psi},\ket{\psi})=\braket{d\psi|d\psi}$. 
To arrive at a local chart , one assumes that the normalized vectors in H smoothly depend on a n-dimensional parameter $\lambda$, to obtain $\ket{\psi}=\sum_\alpha\ket{\partial_\mu\psi(\lambda)}d\lambda^\nu$, with $\partial_\mu\psi =\partial_\psi/\partial\lambda^\mu$.
So the distance becomes $d^2(\ket{\psi}+\ket{d\psi},\ket{\psi})=\sum_{\mu\nu}\braket{\partial_\mu\psi|\partial_\nu\psi}d\lambda^\mu d\lambda^\nu$. Note that we still have the freedom of a phase, $\ket{\psi(\lambda)}\rightarrow e^{i\alpha(\lambda)}\ket{\psi(\lambda)}$.
The question is whether this definition $g_{\mu\nu}=\braket{\partial_\mu\psi|\partial_\nu \psi}-\braket{\partial_\mu\psi|\psi}\braket{\psi |\partial_\nu\psi}$ of the FS metric is sufficient.
Note however, that the FS emerged from the Einstein equations when a cosmological constant was included on the right-hand side. Therefore, we are dealing with a specific physical system.	
This implies, among other things, that the FS metric remains unchanged up to a scalar multiple under the Ricci flow. It also makes $\mathbb{C}P^n$ indispensable to the theory of general relativity, where it serves as a nontrivial solution to the vacuum Einstein field equations.
The FS manifold represents a gravitational instanton, determined by the Einstein equations (note that we did not require a cosmological constant). On the right-hand side of the effective brane manifold, only the  ${\cal E}_{\mu\nu}$ appears.
The K\"ahler form of the FS manifold, when equipped with a real analytic metric, admits a local conformal K\"ahler immersion into a Hilbert space, when a first-order differential equation governing the existence of such an immersion exists. The conformal K\"ahler condition is $d(\Omega^2 K)=0$, where $\Omega$ is a scalar field determined by real physics and $K$ the 2-form. In our case, this will initially be the dilaton field.  Furthermore, even in the case of extreme curvature, we have the freedom of conformal invariance.
In order to say something  about the evolution of distances between states, one  applies  the non-linear Schr\"odinger quation  $-i\hslash d/dt\ket{\psi(t)}=H(t)\ket{\psi(t)}$. 
The path length is then related to the uncertainty in the energy  $\sim 2/\hslash\Delta E t$, with $\Delta E^2=\braket{\psi |H^2|\psi}-\braket{\psi |H|\psi}^2$.
Our model  in K\"ahler compact 2-form, given  by Eq.(\ref{4.23}), could be written in the form as  $d(\omega(r)^2\Omega(r)^2K)=0$, where K is given in Section (3.4).
We could also write our original 'unphysical' manifold in the form of Eq.(\ref{4.1}).
We obtain the desired conformal K\"ahler form, if we choose $\Omega(R)^2=\frac{1}{r(R)^2}$. Note that, regardless of the form of the function $N(r)$, the geometry is conformal K\"ahler.
Next, we must consider the effect of the angular momentum of our manifold, i.e., the $N^{(\varphi)}$. We can make the comparison with the Kerr-de Sitter instanton  metric\cite{page1978}
In this model, there appears also a cosmological constant and  represents a two sphere $S^2$ fibre bundle over a base which is another $S^2$. There are in principle 4 roots of the quartic equation, which can be solved exactly. The two outermost horizons are of importance here. To avoid conical singularities, one must make the $\tau$ variable periodic. Furthermore, the radial coordinate must be redefined using an approximation procedure, as must the azimuthal coordinate.
For no conical singularities at either horizons, both horizons should give the same period and hence the same surface gravity.
Further, the one horizon should rotate around an integral number of times n with respect to the other. However, it is not clear how to make the solution physically acceptable.
The complexification is a better route, just as in the case of the the Schwarzschild to Kerr shift $r\rightarrow r-ia\cos\theta$.
We have in our stationary model only one real singularity, a great difference with Page's model.
We also redefined  the azimuthal angle variable. 
Our angular term decouples completely, so we have a much simpler situation. As said before, we applied the complementarity between the inside and outside observer, because we are interested in the evaporation processes and to describe the primordial instanton. Note that the two different observers use different $\omega$.
To describe the quantum mechanical entanglement of our pair of complex-conjugated instantons in the brane, we will use a different type of FS distance. This differs from the AdS correspondence with a conformal field theory\cite{erdmenger2023}.
If a PBH originates from an instanton pair, its quantum state of entanglement will eventually transition to a classical gravitational state. In this state, pure states are no longer applicable. During the reverse process, i.e. evaporation, the scale decreases over time and the quantum era begins.
Apart from a cosmological term, our bulk space is empty. However, we were able to fine-tune this using the brane cosmological constant. The 'energy' term, ${\cal E}_{\mu\nu},$ appears on the right-hand side of Einstein's equations.
Remember that the dilaton contribution, both in the bulk and on the brane, cannot be considered an additional field. Instead, it is a phantom gravitational field that becomes significant as we approach the Planck scale and interacts with a quantum scalar field. Far beyond the Planck scale, different observers will use different values of omega.
We need a distinguishability criterion for cases where curvature becomes extreme, e.g. black holes. 
The 'distance' between pure states in the associated Hilbert space is equivalent to a complex projected space with an FS metric.
A simplified form is usually considered, where the first 2D block is written as a distinguishability metric on the space of probability distributions,
\begin{equation}
ds^2_{PD}=\sum_j \frac{dp_j^2}{p_j}+\sum_jp_j(d ln p_j)^2=4\sum_j dr_j^2,\label{4.31}
\end{equation}
 with $p_j=r_j^2$. 
Originally, the $p_j$ the probabilities  from a distribution of n samples. One estimates these values as the observed frequencies.
A nearby distribution $\bar p_j$ can be distinguished from $p_j$ if the Gaussian $e^{-(N/2)(\bar p_j-p_j)^2/p_j}$  is small.
Next, a statistical distance is constructed that is generalised to pure states in quantum mechanics. Consider neighbouring pure states expanded using an orthonormal basis  $\ket{j}$, i. e.,   $\ket{\phi}=\sum_j\sqrt{p_j}e^{i\varphi_j}\ket{j}, \ket{\tilde \phi}=\ket{\phi}+\ket{d\phi}=\sum_j\sqrt{p_j+dp_j}e^{i(\varphi_j+d\varphi_j)}\ket{j}$. Normalization then will imply
$Re(\braket{\phi | d\phi})=-1/2 \braket{d\phi | d\phi}$.
Measurements described by the one dimensional projectors $\ket{j}\bra{j}$     can distinguish $\ket{\phi}$ and $\ket{\tilde\phi}$ by the classical metric of Eq.(\ref{4.23}).
The quantum  distinguishability metric  should be defined  by measurements that resolves the two states optimally, i.e., that maximize Eq.(\ref{4.23}).  We already found that the maximum is given by the Hilbert-space angle $\arccos(|\braket{\tilde\phi | \phi}|$, which captures a notion of state  distinguishability. The corresponding line element becomes
\begin{eqnarray}
ds^2_{PS}=(\arccos(|\braket{\tilde\phi |\phi}|)^2=1-|\braket{\tilde\phi |\phi}|^2=\braket{d\phi_\perp|d\phi_\perp}\cr
=\frac{1}{4}\sum_j\frac{dp_j^2}{p_j}+\Bigl[\sum_j p_jd\varphi_j^2-\Bigl(\sum_j p_jd\varphi_j\Bigr)^2\Bigr],\qquad\label{4-32}
\end{eqnarray}
where $\phi_\perp =\ket{d\phi}-\ket{\phi}\braket{\phi |d\phi}$, is the projection of $\ket{d\phi}$ orthogonal to $\ket{\phi}$ and  PS stands for pure state.
The term in brackets is the variance of the phase change. For a suitable basis, this can be set to zero. Therefore, $ds_{PS}^2$ becomes a  maximum.
This means that $ds_{PS}$ is the statistical distance between neighbouring pure states.
This statistical distance can be used to purify quantum states and leads to a Riemannian geometry of density operators. This results in the FS metric for mixed states.
Note that the FS is a gravitational instanton. The GRT solution requires a cosmological constant. Therefore, we are missing the GRT contribution here.
In our case, on the effective brane manifold, we need to project the Weyl tensor from the bulk onto the brane to obtain the same solution. 
In the context of the AdS/CFT model, one could say that our solution on the brane and in the empty bulk (using the RS model) are dual to each other.
In our model, we will partially follow Diósi-Penrose's (DP)\cite{penrose2014} method for forming two different manifolds. However, in order to obtain a single black hole as $a$ increases, we must address the issue in the DP model of one of the instantons dying out.
As we mentioned earlier, this is a gravity-induced approach to wavefunction collapse based on semiclassical gravity, which is enhanced by the hypothesis of a temporally expanding spacetime. This leads to a collapse model that can resolve the conflict between quantum non-locality and relativity.
Let us now return to Eq.(\ref{4.28}), which gives the contribution to the right-hand side of the effective Einstein equations. As in the DP model, each of our instanton states is stationary, as required.
It is not surprising that, in the complex transformed form, the function $\Delta {\cal E}_t^t$ behaves as two sharply located peaks when the complex expression is plotted for given values of   $N^2, \omega$ and $N^{(\varphi)}$ for our model, in the same spots as the complex $N^2$ of Fig.(\ref{fig5}). The constant $C$ determines the 'sharpness' of the smeared-out delta-like function. 
Starting from an initial situation of a very small scale $a$, which is related to the mass, we obtain two different complex manifolds. As $a$ increases, the separation increases and the collapse will occur faster. As in the DP model, we can estimate the expected time by applying Heisenberg uncertainty: $\Delta\tau\sim\hslash/\Delta {\cal E}_t^t$, where $\Delta {\cal E}_t^t$ represents the uncertainty in the energy of the superposed state.
In the DP model, in some way, one spacetime must survive due to an instability in one branch.
\begin{figure}[h]
\centerline{
\includegraphics[width=0.3\textwidth]{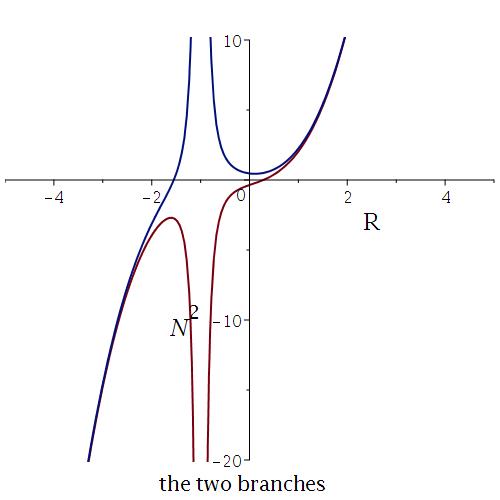}}
\caption{{\it Plot of the two branches for the real coordinate transformation $r=R+2a$. We took for $C=\pm \frac{3381 a^5}{256}$ and $a=1$.}}\label{N2real}
\end{figure}
In Fig.(\ref{N2real}) we plotted two branches, in the case of the real coordinate transformation, comparable with the EH instanton, in order to make it geodesically complete and free of true singularities. For the complex coordinate transformation we obtained  two different intrinsic solutions of $ N^2$, a pair of instantons in the complex plane. It is remarkable that the behavior is insensitive for the sign of $C$.
\begin{figure}[h]
\centerline{
\includegraphics[width=0.26\textwidth]{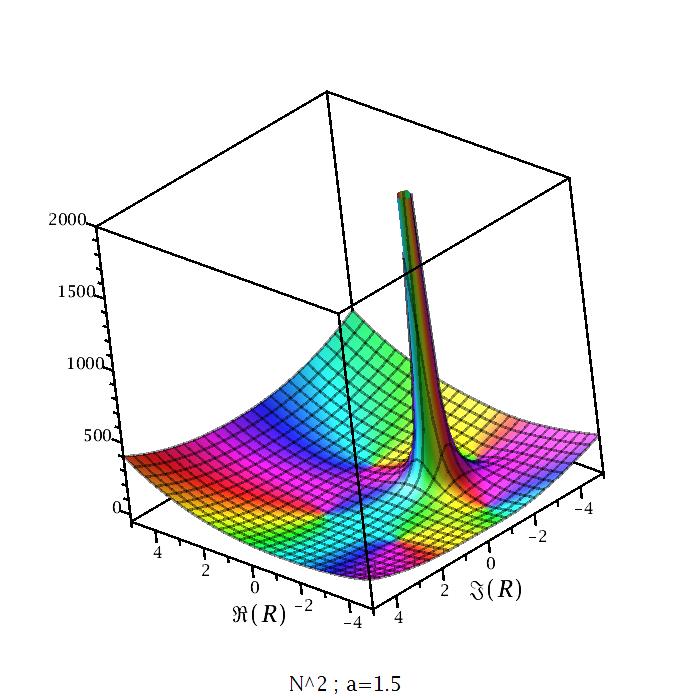}
\includegraphics[width=0.26\textwidth]{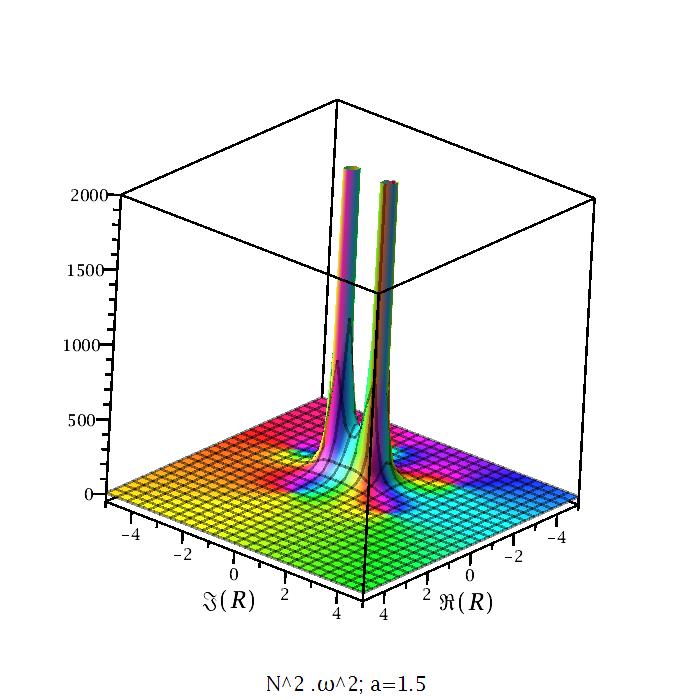}}
\caption{{\it  Plot of $N^2$ and $\omega^2 N^2$ of the plus-branch  for the complex coordinate transformation $r=a\pm i(R+a) $. We took for $C=\pm \frac{3381 a^5}{256}$ the plus sign and $a=1.5$. Remarkably, for the 'unphysical' metric, we have a single instanton, whereas for the 4D brane metric, there appears a double peak when the metric includes the term  $\omega^2$. }}\label{N2compom}
\end{figure}
We conjecture that we require both complex solutions. If $a$ increases from the Planck era to the GRT era and back to real coordinates, we will apply the antipodal map instead of handling region III of the Penrose diagram, where some people suggest a white hole sits. The Klein surface embedding possesses the $\mathbb{Z}_2$  symmetry. In 5D, we are dealing with two copies of the Klein bottle. Through fibering, we obtained the $S^3$, the 4D effective brane.
For an observer outside the brane, there are two $S^3\otimes \mathbb{R}$. As the instanton solution follows from both manifolds, there are no obstructions when we introduce the real primordial black hole into the brane. 
In Fig.(\ref{N2compom}) we plotted for the complex case  $N^2$, i.e., the 'un-physical' metric, and $\omega^2 N^2$ for $C= \frac{3381 a^5}{256}$ and $a=1$. We took the plus-branch. We observe that the double solution reappears for our effective 4D brane metric. When $a$ decreases and one approachs the very small scale, $\omega$ becomes a quantum field along with a possible scalar field.
Let us be more specific about wave function collapse. The state vector evolves abruptly at critical spacetime expansions, changing to a trajectory that resembles a classical one. This can explain the correlations in EPR experiments without contradicting relativity, since the evolution of the state vector before and after the abrupt change is governed solely by local physical laws. It would be interesting to obtain lifetimes of superpositions using the gravity-based approaches of the DP model. In our case of the PBH, it must be realized that the time evolution can be described by our parameters $(a, C)$.
The goal, of course, is to overcome the conflict between quantum non-locality and relativity.
The evolution of the state vector of the two states with energy distribution ${\cal E}_t^t(x_1)$ and ${\cal E}_t^t(x_2)$, emerges just from the initial singularity in the complex plane.
The projected Weyl tensor contribution evolves on our brane world. This represents a 'traceless energy density' on the right-hand side of the 4D Einstein equations.
The evolution of the state vector of the two states  ${\cal E}_t^t(x_1)$ and ${\cal E}_t^t(x_2)$ can be written as $\ket{\phi}=c_1\ket{\phi_1}+c_2\ket{\phi_2}$. The mass distribution can then we written as ${\cal E}_t^t(x)=|c_1|^2{\cal E}_t^t(x_1) +|c_2|^2{\cal E}_t^t(x_2)  $,   with $ {\cal E}_t^t(x) =\braket{\phi_i| \hat  {\cal E}_t^t(x)|\phi_i|}$.          
The common gravitational potential $V_c(x)$ is then $V_c(x)=|c_1|^2 V_1(x)+|c_2|^2V_2(x)$.\\
\subsection{Collapse of the wave function: alternatives? }
Physicists are still grappling with the complex concepts of indeterminacy and non-locality in quantum mechanics, whether or not combined with general relativity, as in  Penrose's model.
There are physicists who have doubts about the Copenhagen interpretation of quantum mechanics:  {\it 'The Copenhagen doctrine seems to be the only way to describe what is going on in Hilbert space controlled by operator equations. However, the description of our classical world will be forever impossible, unless one accepts some extremely contrived theoretical contraptions that may or not reproduce the QM as observed in experiments'}. Bell?s theorem would be resolved in one fell swoop, if one were to abandon the idea of free will and accept a form of super-determinism.  
A proposition of super-determinism:  'Probabilistic systems are in reality, deterministic systems in disguise'. In a truly deterministic world, there can be, in principle, exist correlations between particles and the choices 'A' and 'B'. In these models, it is assumed  that all events have causes tracing back to the Big Bang.
There are many attempts to restore a more classic character to QM, notably by efforts aiming at deriving the theory from a more fundamental, deeper-lying theory. To name just a few, the 'pilot-wave' system and 't Hooft's 'cellular interpretation of QM'. 
The most pressing question, of course, remains whether gravity can play a role in situations involving extreme curvature.
Let us return to Penrose's model, where a spatial quantum superposition collapses as a back-reaction from spacetime, which is curved in different ways by each branch of the superposition.
One can say that there is a clash of basic principles, i. e., 'principle of equivalence and general covariance' and 'superposition principle'.
Penrose on 'gravitization' of QM: {\it 'Nature 'dislikes' and therefore suppresses superposition of  locally different spacetime curvatures'.}
The more massive the system in the superposition, the greater the difference between the two spacetimes and the faster the wave function will collapse. 
The expected time is $\Delta\tau\sim \hslash/\Delta E_{G}$, where  $\Delta E_G $ measures how large the superposition is (the gravitational self-energy of difference between the  mass distribution). $\Delta E_G$ can be seen as a fundamental uncertainty in the energy of the superposed state and follows from Heisenberg's uncertainty relation.  The instability originates from some non-linear effects due to the  gravitization' with lifetime $\tau\sim \hslash/E_G$. 
One considers a system with mass density 
\begin{equation}
\mathcal{M}(x)=\Sigma_j m_j\mu_{R_0}(x-\hat x_j),\label{4-33}
\end{equation}
with $ m_j, \hat x_j,$ and $ \mu_{R_0}$ the mass, position operator and mass density function of the j-th particle, respectively. $R_0$ is a parameter to smear the mass density function.
This approach has some subtle difference with the standard 'conscious perception' of quantum formalism.
The question is how to modify the Schr\"odinger equation. The DP model suggests
\begin{equation}
i\hslash\frac{\partial\Psi}{\partial t}=-\frac{\hslash^2}{2m}\nabla^2\Psi+V\Psi,\label{4-34}
\end{equation}
where the potential V represents the gravitational self-energy. When the separation of the lump of mass becomes significant in a quantum-gravity sense, then one branch dies off and the other survives.  The instability that results in just one of the spacetime branches  surviving set, happens at the Planck scale.
In the  previous section we made a connection between our model and the DP model by  switching  to the complex space. There is yet  another connection.
A remarkable paper has recently been published on an asymmetric intrinsic non-Gaussian longitudinal momentum transport distribution of heavy quark particles in a quark-gluon plasma in the very early universe\cite{plessis2026}. In the first fraction of a second following the Big Bang, matter existed as a trillion-degree plasma of quarks and gluons, that zinged around at close to the speed of light. This roiling soup lasted for just a few microseconds, before the particles coalesced into more stable protons and neutrons. In some sense, one can compare this phenomenon with the Brownian motion.
This model picks up where Einstein's Brownian movement model left off, to predict how particles move through fluid under extreme conditions. 
What was done is tracking other particles, such as heavy quarks, that are not part of the plasma but happen to move through it. The way a heavy particle moves through a fluid should tell something about the microscopic structure of the fluid itself.   
The expected Gaussian distribution of the random kicks will break down. The authors assumes that the strong force is responsible for the binding of the quarks and gluons.
They found a imbalance in the bell-shaped distribution. The tails at either ends of the curve were uneven. The large kicks become rare.
A weak point is that one cannot ignore gravity in this very early era of the universe, just as in the case of PBH models. What if we could consider  the instanton pair as the 'heavy particles' in the spacetime foam, in stead of the heavy quarks? One must assume then that there is already a complex FS manifold. 
Finally we mention an extraordinary study, a long time ago by Wiener\cite{wiener1958}. Let us consider a complex differentiable space with  $Y(\xi, \alpha$) a complex expression representing, a formal Fourier coefficient of, for example, a Brownian motion. 
One writes
\begin{equation}
\int_{-\infty}^{\infty}\phi_n(\xi)dY(\xi,\alpha)=A_n(\alpha)+i B_n(\beta),\label{4-35}
\end{equation}
where A and B have Gaussian distributions. For $\phi_n$'s orthogonal, the distributions are independent.
Let us consider the distribution in polar coordinates in the complex plane
\begin{equation}
P(r,\theta)=\frac{1}{2\pi} e^{-r^2/2}r drd\theta.\label{4-36}
\end{equation}
Integrating over $\theta$, we obtain
\begin{equation}
P(r)dr=r e^{-r^2/2}dr. \label{4-37}
\end{equation}
Next one considers two distribution expressions
\begin{eqnarray}
r_1=\int\phi_1(\xi)dY(\xi,\alpha)=A_1(\alpha)+iB_1(\alpha),\cr r_2=\int\phi_2(\xi)dY(\xi,\alpha)=A_2(\alpha)+iB_2(\alpha).\label{4-38}
\end{eqnarray}
\begin{figure}[h]
\centerline{
\fbox{\includegraphics[width=0.23\textwidth]{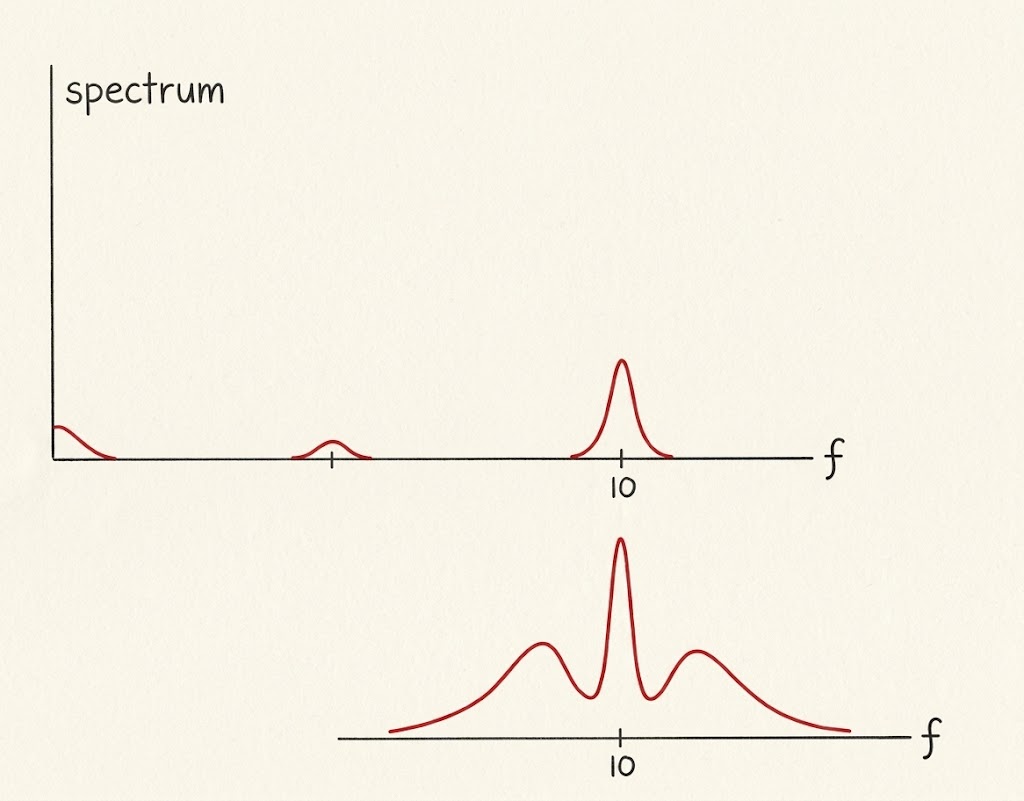}}
\fbox{\includegraphics[width=0.21\textwidth]{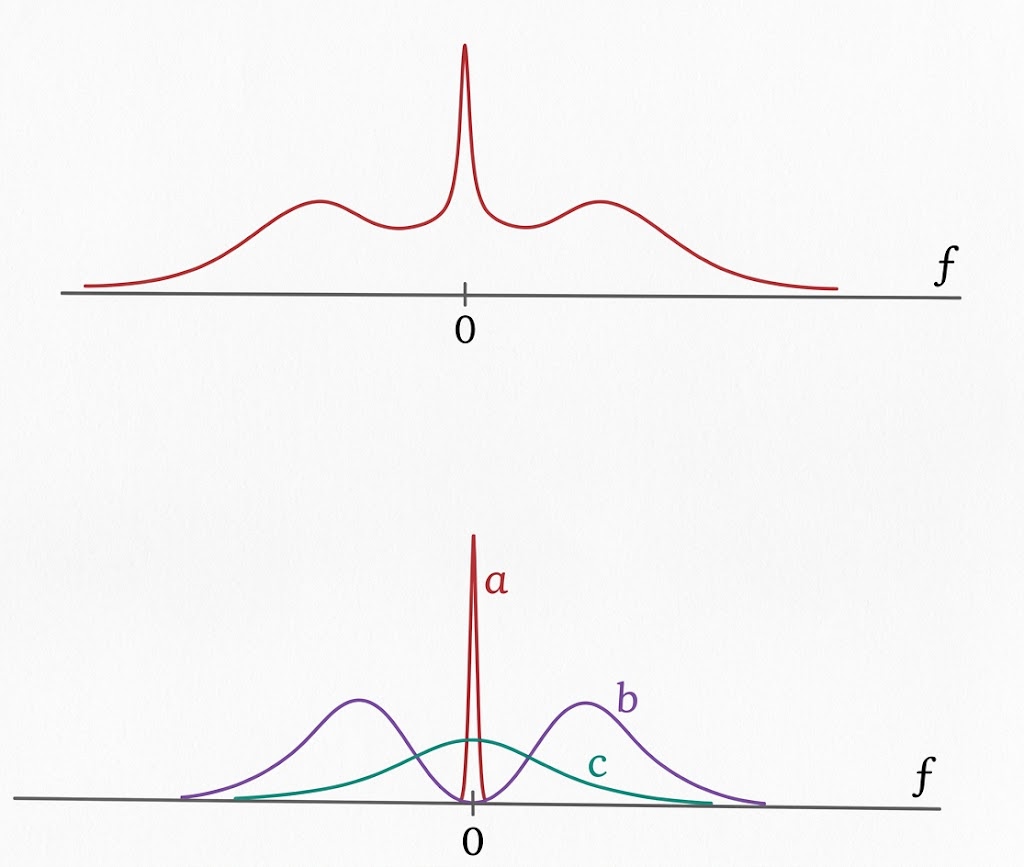}}}
\caption{{\it Left, top: Spectrum of the wave brain, with the $\alpha$ rhythm at 10 Hz. Left, below: A high resolution pulse around 10 Hz. A 'pulling-together' interaction effect by non-linearities.  Right, top: the shifted spectrum to the origin. See text. Right, bottom: The leading parts of the spectrum. This 'dip' phenomenon will occur in many physical situations\cite{wiener1958}.}}\label{wiener}
\end{figure}
One  checks that
\begin{equation}
P[|a|r_1 >|b|r_2]=\frac{|a|^2}{|a|^2+|b|^2}.\label{4-39}
\end{equation}
In general, one can extend  a randomness function $X(t,\alpha)$ to higher dimensions. 
The variable t (usually time) can then be replaced by an s-dimensional space. One has again two independent Gaussian distributions, due to orthogonality.
$X(s,\alpha)$ defines  an additive, but not absolute, function of measurable sets $\int T_i(s)dx(s,\alpha)$. Here $T_i(s)$ is a function which is 1 on $T_i$ and 0 elsewhere.
If one takes, for example, the  plane, one divides it in 4 equal segments of $[0,1]$. One repeat this procedure for every segment, to obtain 16 squares, etc. This delivers  a measure-preserving transformation between square measure on the plane and linear measure on the line. This method will go over to higher dimensions.  We have s dimensions and a random additive functional. The most interesting case is that of complex numbers, which we have already discussed.
$X(s,\alpha)$ represents now a  measure  preserving mapping, where the dimension is not relevant. We know that in QM a state at one particular time is $\sum a_n\phi_n(s)$, with $\sum|a_n|^2=1$.
The Schr\"odinger equation then determines the time evolution, $\sum a_ne^{i\lambda_nt}\phi_n(s)$, with $|a_n|^2$ the probability, $\phi_n$ eigenfunctions and $a_n$ the characteristic amplitudes. The probability is not variable with time. The state is stationary. You could say that Schr\"odinger equation  can be considered  as determining a group of measure preserving transformations in space.
If there is no measurement, the state is in superposition. Could this be the case for our model?
Let us return to two eigenfunctions with probabilities $|a_1|^2/(|a_1|^2+|a_2|^2)$ and  $ |a_2|^2/(|a_1|^2+|a_2|^2)$, 
belonging to $a_1\phi_1(s)$ and $a_2\phi_2(s)$. They generate
\begin{equation}
A=a_1\int\phi_1(s)dy(s,\alpha),\quad B=a_2\int\phi_2(s)dy(s,\alpha).\label{4-40}
\end{equation}
Again we can write
\begin{equation}
P(|A|\geq|B|)=\frac{|a_1|^2}{|a_1|^2+|a_2|^2}.\label{4-41}
\end{equation}
It is equals a Lebesgue measure. If the amplitude $a_1$ grows by a non-linear instability (by the expansion parameter $\epsilon$) one of the eigenfunctions will 'survive', which in QM means a choice between 2 eigenfunctions. 
Now we can place this results in relation with Fig.(\ref{wiener}). Could this be a 'collapse' of the wave function?

An different alternative for the description of stationary states in super position, was delivered by Bialynicki-Birula\cite{bial1976}. They consider a modified Schr\"odinger equation with non-linear terms, such that the solution possesses localized stationary solutions for the state of macro-objects, 
\begin{equation}
i\hslash\frac{\partial}{\partial r}\psi(r_k,t)=\Bigl(-\frac{\hslash^2}{2m}\nabla+V(r_k,t)+F_k(|\psi_k|^2\Bigr)\psi_k(r_k,t),\label{4-42}
\end{equation}
where $r_k$ ($k=1,2$) belong to two separable non-interacting subsystems $\psi(r_1,t)$ and $\psi(r_2,t)$ and $F_k(|\psi_k|^2$ the non-linear term such as $\sim \psi\log\psi^2$.
The question is if  the superposed state is still stationary? QM predicts non-stationary behavior and in principle detectable. 

It is remarkable, that the equation is invariant under rotation of spin.
If we have two noninteracting subsystems, $U=U=U_1(x_i,t) +U_2(x_i,t)$, then we can write a solution of the Schr\"odinger equation in the form of a product of two solutions subsystems. So if subsystems are initially uncorrelated, they will remain uncorrelated. Correlations can be introduced only by mutual interactions.
However, if we allow a nonlinear term in the Schr\"odinger equation,  then, in general, it will produce correlations. If we consider a system of two particles, the very existence of a second particle influences the spreading of the wave packet of the first particle. Finally one proves that the only nonlinear equation which obeys the separability postulate, is
\begin{equation}
i\hslash\frac{\partial}{\partial t}\psi(r,t)=\Bigl(-\frac{\hslash^2}{2m}\nabla+U(r,t)-b\ln(|\psi|^2a^n)\Bigr)\psi(r,t)\label{4-43}
\end{equation}
with a a constant with dimension of length and n the dimension of configuration space. When $\psi(r,t)$ is a solution, so is $\alpha\psi(r,t)e^{i/\hslash bt\ln|\alpha|^2}$.
One assumes $\int d^nr\psi^*(r)\psi(r)=1$. Further, it conserves the property of  separation of variables, e. g., the center of mass motion for an isolated system.
Further, the energy-momentum tensor will contain a  logarithmic nonlinear term $\sim \nabla_i\nabla_j\ln |\psi|^2$.
A shortcoming of this type of model is that the curvature-induced distortion of the background is not taken into account, or is only accounted for in an approximate form, $g_{\mu\nu}=\bar g_{\mu\nu}+\epsilon h_{\mu\nu}$, with an rather unclear expansion parameter. 

An interesting method, suitable in handling this back reaction in this context, is the application of the high-frequency, or multiple-scale, method of Choquet-Bruhat\cite{choq1969,slagter1986}. It can handle in a covariant way, the back-reaction  on the metric. The expansion parameter can be taken, as an example, the ratio of the background scale to the fast varying scale, which often occurs in critical situations. In our case, it could be the the ratio of bulk scale to the brane scale.
Moreover, the method also delivers the non-linear Schr\"odinger equation from the well known $\Phi^4$ model, $\Phi_{tt}-\Phi_{xx}=-(\beta\Phi^2-\alpha)\Phi$\cite{slagter2025b}.
The method is based on developing the relevant fields $V_i$ as
\begin{equation}
V_i=\sum_{n=0}^{\infty}\frac{1}{\epsilon^n}F_i^{(n)}({\bf x},\xi),\label{4-44}
\end{equation}
where $\epsilon$ represents again a dimensionless  parameter ('frequency'), which will be large. Furthermore, $\xi=\epsilon\Theta({\bf x})$, with $\Theta$ a scalar (phase) function on the manifold.

The interested reader is also referred to the recent work of Adler et al.\cite{adler2015,adler2016}).  This model posits that 'noise', a prerequisite for theories of objective state-vector reduction, can originate from fluctuations in the complex aspect of spacetime. State-vector reduction is driven by a 'spacetime foam' that is valued in complex numbers. 
The metric is written as the sum of a real and a fluctuating complex spacetime: $g_{\mu\nu}=\bar g_{\mu\nu}+h_{\mu\nu}$. 
The method employs the principles of state vector normalisation and the prohibition of faster-than-light signaling. This results in the usual de-correlation function.
The author also identifies a square root branch cut in polar coordinates, beginning at r = a, which becomes complex below a finite radius. 
This method can also be interpreted as a Hermitian Kähler form.
However, even in his model, it is unclear how the back reaction can be consistently treated in the case of strong curvature.

Clearly, these connections need to be followed up. The author is currently studying this.
\section{Summary}\label{sec:5}
\begin{figure}[h]
\centerline{	
	\includegraphics[width=8.cm]{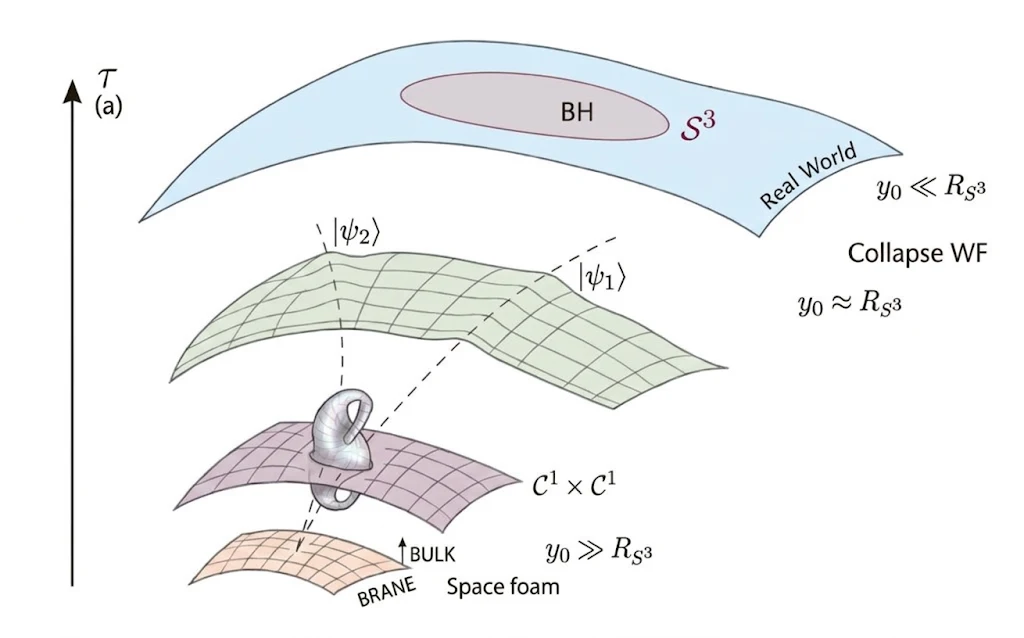}}
	\caption{ Our model in a nutshell}		
	\label{summ}
\end{figure}

In a Randall-Sundrum-type brane world model, we found an exact instanton solution that satisfies both the 5D and the effective 4D field equations on a Kerr-like stationary manifold.
The dilaton, considered as a quantum field will, in general, interact with a complex scalar field, which is believed to have been present during that era. A full-scale quantum gravity model will be required. Here, we initially considered a vacuum situation, with only brane and bulk stresses. The constraint equations can be met by fine-tuning these constants.
As the brane scale grows (i.e., $a$), we must initially transfer the coordinate system to the complex plane in order to manage the singularities. $a$ ($R_{S^3}$ of our $S^3$) will be much smaller than $y_0$.
Two entangled instantons seem to emerge, and Penrose's insights are applicable: it is gravity that will also contribute to collapsing the quantum system. Our set of equations is fully covariant.
Fig. (\ref{summ}) provides a pictorial representation. We also use the full Fubini-Study metric, a conformal K\"ahlerian instanton manifold, to describe the distance between the correlated states, rather than using a flat replacement. The Fubini-Study manifold requires a 'matter' term on the right-hand side of Einstein's equations. In our case, this is the contribution from the five-dimensional Weyl tensor, also known as the Kaluza?Klein contribution.
As $a$ increases, the wave function will collapse when $a\sim y_0$ . Finally, when $y_0 << R_{S^3}$ we can describe our black hole using the effective brane manifold. At this point, the contribution from the bulk to the brane is required to ensure consistency with the bulk Einstein equations.
Similarly, we could describe the evaporation of the black hole by Hawking radiation using the reverse treatment, because we have CPT invariance.
Furthermore, we could apply the antipodal map such that the radiation remains pure.  We need also the topology of the Klein surface with  $\mathbb{Z}_2$ symmetry to describe the fibering from 5D to our $S^3$. The non-orientable Klein bottle replaces the non-orientable M\"obius band in conventional antipodicity models.
We related our solution to the quantum geometry of the icosahedron. We found a remarkable connection between the zeros of the relevant quintic polynomial and the fractional Möbius transformations. Finally, we discovered a potential connection between the distinguishing distance of two probability distributions and the instanton pair's Riemannian manifold.

\end{document}